%% file: CCPE.tex
\documentclass[12pt,a4]{article}
\usepackage{color}
\usepackage{multirow}
\usepackage{mathptm}
\usepackage{amsfonts}
\usepackage{amsmath}
\usepackage{amsthm}
\usepackage{amssymb}
\usepackage{graphicx}
\usepackage{sectsty}
\usepackage{bbold}
\usepackage{subfigure}
\usepackage{algorithm}
\usepackage{algorithmicx}
\usepackage{algpseudocode}
\usepackage{url}
\algrenewcommand{\alglinenumber}[1]{\footnotesize#1}

\input{symacros}
\begin{document}
%
%
\title{An Optimized and Scalable Eigensolver for Sequences of
  Eigenvalue Problems\footnote{Article based on research supported by
    the Excellence Initiative of the German federal and
    state governments and the J\"ulich Aachen Research Alliance –
    High-Performance Computing.}}

\author{%
  Mario Berljafa \footnote{School of Mathematics, The University of
    Manchester, Alan Turing Building, M13 9PL, Manchester, United
    Kingdom. {\tt m.berljafa@maths.man.ac.uk}.}  \and Daniel Wortmann
  \footnote{Institute for Advanced Simulation and Peter Gr\"unberg
    Institut, Forschungszentrum J\"ulich and JARA, 52425 J\"ulich,
    Germany. {\tt d.wortmann@fz-juelich.de}.} \and Edoardo Di
  Napoli
  \footnote{J\"ulich Supercomputing Centre, Institute for Advanced
    Simulation, Forschungszentrum J\"ulich and JARA, Wilhelm-Johnen
    strasse, 52425 J\"ulich, Germany.  {\tt
      e.di.napoli@fz-juelich.de}.} }
\maketitle

\begin{abstract}
  In many scientific applications the solution of non-linear
  differential equations are obtained through the set-up and solution
  of a number of successive eigenproblems. These eigenproblems can be
  regarded as a sequence whenever the solution of one problem fosters
  the initialization of the next. In addition, in some eigenproblem
  sequences there is a connection between the solutions of adjacent
  eigenproblems. Whenever it is possible to unravel the existence of
  such a connection, the eigenproblem sequence is said to be
  correlated.  When facing with a sequence of correlated eigenproblems
  the current strategy amounts to solving each eigenproblem in
  isolation. We propose a alternative approach which exploits such
  correlation through the use of an eigensolver based on subspace
  iteration and accelerated with Chebyshev polynomials (ChFSI). The
  resulting eigensolver is optimized by minimizing the number of
  matrix-vector multiplications and parallelized using the Elemental
  library framework. Numerical results show that ChFSI achieves
  excellent scalability and is competitive with current dense linear
  algebra parallel eigensolvers.
\end{abstract}

\newpage
\pagenumbering{arabic}
%
%
%
\section{Introduction}
\label{sec:intro}

In many scientific applications the solution of Hermitian
eigenproblems is key to a successful numerical simulation. A
considerable subset of applications requires the solution of not just
one eigenproblem but a sequence of them. An exemplary case is provided
by the non-linear eigenvalue problem where the solution is computed
through a step-wise self-consistent process: at each step the solution
of a linearized eigenvalue problem is used to initialize the next one
until an application-specific criterion is satisfied. In this context a
well-known example is represented by Density Functional Theory (DFT),
one of the most important frameworks used in Material Science and
Quantum Chemistry.

While there is a large amount of literature on algorithms developed to
solve for an isolated Hermitian (standard or generalized) eigenvalue
problem, the same cannot be said for a sequence of them. It is the aim
of this paper to illustrate an alternative approach to solve a
sequence of Hermitian eigenproblems when a degree of correlation
between their eigenpairs is made manifest. We take inspiration
from sequences of dense eigenproblems arising in DFT, but want to
stress that such approach can work as well in other applications
provided that the sequence possesses analogous properties.

{\bf Eigenproblem sequences ---} It is important to distinguish
between the notion of sequence of eigenproblems and the correlation
that might exist between adjacent eigenproblems in a sequence. The
concept of sequence is very general and could be loosely borrowed
from how sequences of real numbers are defined.
\begin{definition}
  A sequence of eigenproblems is defined as an $N$-tuple $\{P\}_N
  \equiv P^{(1)}, \dots, P^{(\ell)}, \dots, P^{(N)}$ of problems
  $P^{(\ell)}$ with same size $n$ such that the eigenpairs of the
  $\ell$-problem are used (directly or indirectly) to initialize the
  $\ell+1$-problem.
\end{definition}
\noindent
The above definition does not, in general, imply that a connection
between adjacent problems in a sequence is direct or simple. It just
states that the sequence is generated through a output/input process:
the solutions of an eigenproblem in the sequence are manipulated to
generate the next eigenproblem. In order to express more precisely the
idea of ``connection'' between eigenproblems, we introduced the notion
of correlation (not to be confused to the definition of correlation
used in probability calculus).
\begin{definition}
  Two eigenproblems $P^{(\ell)}$ and $P^{(\ell+1)}$ are said to be
  correlated when the eigenpairs of the $\ell+1$-problem are in some
  relation with the eigenpairs of the $\ell$-problem.
\end{definition}
\noindent
There may be many ways to measure a correlation between eigenpairs of
two distinct eigenproblems. For instance, the eigenvalues of one
matrix could be shifted with respect to the eigenvalues of the
previous one by a value proportional to the eigenvalue itself. In this
paper we address correlation between eigenpairs by measuring the
angle between corresponding eigenvectors as a function of the sequence
index $\ell$. Whenever it is difficult to find such a correlation in
exact mathematics a numerical approach may be
preferred~\cite{DiNapoli:2012fk}.

The current approach to solve for eigenproblem sequences depends on
the properties of the matrices defining the problems and on the
percentage of eigenspectrum required. Unless very few eigenpairs are
required, dense eigenproblems are solved with direct eigensolvers
provided by libraries such as LAPACK~\cite{Anderson:1999vp}. Sparse
eigenproblems prefer iterative eigensolvers, the most popular of which
is provided by the ARPACK package~\cite{lehoucq1998arpack}. In both cases the
invoked eigensolvers are used as black-boxes and the correlation
between eigenproblems (when present) is not exploited. Moreover, it
was recently shown that when used on parallel architectures iterative
eigensolvers can still be used on dense problems and be competitive
with direct ones~\cite{Aliaga:2012bx}. Together, these last two
observations, indicate the need for a viable third option specifically
dedicated to sequences of eigenproblems.

In this paper we illustrate such an alternative approach to solve
sequences which show some degree of eigenvector correlation. The
novelty of the approach consists of developing a method which takes
advantage of the available information provided by the correlation. To
be specific, we exploit the eigenvectors (and the extremal
eigenvalues) of the $\ell$-problem in order to facilitate the
computation of the solution of $\ell+1$-problem. This is made possible
by inputting the $\ell$-eigenvectors to a simple block eigensolver
based on subspace iteration accelerated with Chebyshev polynomials
~\cite{Rutishauser:1969ub,Zhou:2006ek}. We show that such a block
eigensolver can be further optimized by appropriately tuning the
polynomial degree of the accelerator. When parallelized on distributed
memory architectures our Chebyshev Filtered Subspace Iteration (ChFSI)
is a viable alternative to conventional eigensolvers both in terms of
scalability and performance.

{\bf Sequences in DFT ---} Within the realm of
condensed-matter physics, DFT is considered the standard model to run
accurate simulations of materials. The importance of these simulations
is two-fold. On the one hand they are used to verify the correctness
of the quantum mechanical interpretation of existing materials. On the
other hand, simulations constitute an extraordinary tool to verify the
validity of new atomistic models which may ultimately lead to the
invention of brand new materials.

Each simulation consists of a series of self-consistent field (SCF)
cycles; within each cycle a fixed number $\mathcal{N}_{\ \kv}$ of
independent eigenvalue problems is solved. Since dozens of cycles are
necessary to complete one simulation, one ends up with $\mathcal{N}_{\
  \kv}$ sequences made of dozens of eigenproblems. The properties of
these eigenproblems depend on the discretization strategy of the
specific DFT method of choice. In this paper we will exclusively
consider the Full-Potential Linearized Augmented Plane Waves method
(FLAPW). FLAPW gives rise to sequences of dense hermitian generalized
eigenproblems (GHEVP) with matrix size typically ranging from 2,000 to
30,000. In FLAPW only a fraction of the lowest part of the eigenspectrum is
required. The eigenvalues inside this fraction correspond to the
energy levels below Fermi energy and their number never falls below
2\% or exceeds 20\% of the eigenspectrum. The relatively high number
of eigenpairs in combination with the dense nature and the size of the
eigenproblems inevitably lead to the choice of direct
eigensolvers.

Until very recently, the computational strategy on parallel
distributed memory architecture favored the use of
ScaLAPACK~\cite{Blackford:1987ta} implementation of the bisection and
inverse iteration algorithm (BXINV). Modern and efficient dense
libraries, like ELPA~\cite{Auckenthaler:2011cy} and
EleMRRR~\cite{Petschow:2013vc}, improve the performance but do not
change the overall computational strategy: each problem in the
sequence is solved in complete independence from the previous one. The
latter choice is based on the view that problems in the sequence are
considered only loosely connected. In fact despite the solution of one
problem ultimately determines the initialization of the next, it does
so in such a mathematically indirect manner that the solutions of two
successive problems seem to be quite independent.

Due to their properties and the current strategy used to solve them,
the FLAPW generated sequences constitute the ideal ground to
illustrate the workings for the ChFSI eigensolver. This is made
possible because, in spite of the assumed loose connection between
adjacent eigenproblems, it has been shown that they possess a good
degree of correlation which is explicitly expressed by the evolution
of the eigenvectors angles as a function of the sequence index
$\ell$~\cite{DiNapoli:2012fk}.

This paper is an invited contribution and it has been specifically
written as an extension of an earlier
publication~\cite{BerDin:LNCS}. In particular we have substantially
enlarged each section with new material, included new numerical tests
and plots and devoted an entire subsection to the optimization of the
polynomial filter. In Sec.~\ref{sec:SeqDens} we illustrate how
sequences of eigenproblems arise in Density Functional Theory and
briefly present results on correlated sequences. The ChFSI algorithm
is described in Sec.~\ref{sec:ChFSI}. Here we give details on the
algorithm structure with special focus on the Chebyshev filter. We
also devote a subsection to the MPI parallelization of the whole
code. In Sec.~\ref{sec:numtest} we present a number of numerical tests
and their analysis. We summarize and conclude in
Sec.~\ref{sec:sumconc}.


  \section{Sequences of correlated eigenproblems}
  \label{sec:SeqDens}

  In this section we describe how sequences of GHEVP arise in
  electronic structure computations. In the first part of the section,
  we provide the reader with an outline of the mathematical set-up and
  illustrate some of the important features which justify interpreting
  the generated eigenproblems as a sequence. In the second part, we
  show how these sequences are also correlated and how such
  correlation can be harnessed and exploited in order to speed-up the
  solution of the whole sequence of problems. We want to stress once
  again that the particular application where these sequences appear
  constitute just a practical example of our method. Such a method
  can be applied to any sequence of eigenproblems endowed with a
  reasonable level of correlation between its eigenvectors.


  \subsection{Eigenproblems in Density Functional Theory}
  \label{sec:seqDFDT}

Every DFT method is based on a variational principle stemming from the
fundamental work of Kohn and Hohenberg~\cite{Hohenberg:1964fz}, and its
practical realization~\cite{Kohn:1965zzb}. Central to DFT is the
solution of a large number of coupled one-particle Schr\"odinger-like
equations known as Kohn-Sham (KS).
\[
\Ham_{\rm KS}\phi_a({\bf r}) \equiv \left( \frac{\hbar^2}{2m}\nabla^2
+ \mathcal{V}_{\rm eff}[\nr] \right) \phi_a(\rv) = \epsilon_a \phi_a(\rv)
\quad ; \quad \nr = \sum_a f_a |\phi_a(\rv)|^2
\]
Due to the dependence of the effective potential $\mathcal{V}_{\rm
  eff}$ on the charge density $\nr$, in itself a function of the
orbital wave functions $\phi_a(\rv)$, the KS equations are non-linear
and are generally solved through a self-consistent process.
Schematically, such process is divided in a sequence of
outer-iterative SCF cycles: it starts off with an initial charge density
$n_{\rm init}({\bf r})$, proceeds through a series of iterations and
converges to a final density $n^{(N)}({\bf r})$ such that $|n^{(N)} -
n^{(N-1)}| < \eta$, with $\eta$ as an a priori parameter.
\\[5mm]
\be
  \label{eq:cycle}
	\begin{array}[l]{ccccc}
          \color{red}{\fbox{\rule[-0.2cm]{0cm}{0.6cm}\shortstack{{\small Initial input}\\$n_{\rm init}(\rv)$}}} &
          \Longrightarrow & \fbox{\rule[-0.2cm]{0cm}{0.6cm} \shortstack{{\small Compute KS potential}\\$\mathcal{V}_{\rm ef{}f}[\nr]$}} &
          \longrightarrow & \fbox{\rule[-0.2cm]{0cm}{0.6cm} \shortstack{{\small Solve KS equations}\\$\Ham_{\rm KS}\phi_a({\bf r}) = \epsilon_a \phi_a({\bf r})$}} \\[4mm]
          & & \uparrow {\rm No} & & \downarrow  \\[2mm]
          \color{blue}{\fbox{\rule[-0.2cm]{0cm}{0.6cm} \shortstack{{\small OUTPUT}\\ {\small Energy, forces, etc.}}}} &
          \shortstack{{\small Yes} \\ $\Longleftarrow$} &  \fbox{\rule[-0.2cm]{0cm}{0.6cm} \shortstack{\small Converged? \\ $|n^{(\ell)} - n^{(\ell-1)}| < \eta$}} &
          \longleftarrow &  \fbox{\rule[-0.2cm]{0cm}{0.6cm} \shortstack{{\small Compute new density}\\$n(\rv) = \sum_a f_a |\phi_a({\bf r})|^2$}}
 	\end{array}
\ee
\\[5mm]
In practice this outer-iterative cycle is still quite computationally
challenging and requires some form of broadly defined discretization.
Intended in its broadest numerical sense, the discretization
translates the KS equations in a non-linear eigenvalue
problem.

Eigenproblems generated by distinct discretization schemes
have numerical properties that are often significantly different; for
sake of simplicity we can group most of the schemes in three
classes. The first and the second classes make respectively use of
plane waves and localized functions to expand the one-particle orbital
wave functions $\phi_a(\rv)$ appearing in the KS equations
\begin{equation}
\label{eq:linearcomb}
\phi_a(\rv) \longrightarrow \phi_{\kv,i}(\rv) =
\sum_{\bf G} c^{\bf G}_{\kv,i} \psi_{\bf G}(\kv,\rv).
\end{equation}
Methods in the third class do not use an explicit basis for the
$\phi_a(\rv)$'s but discretize the KS equations on a grid in real
space using finite differences. The eigenvalue problems emerging from
the first two discretization classes consist of dense matrices of
small-to-moderate size while, within real space methods, one ends up
with very large sparse matrices. Due to the dramatically different set
of properties of the eigenproblems, each DFT method uses a distinct
strategy in solving for the required eigenpairs. For instance it is
quite common that methods based on plane waves use direct eigensolvers
while real space methods make use of iterative eigensolver based on
Krylov- or Davidson-like subspace construction. From the point of view
of software packages for distributed memory architectures, the choice
between direct or iterative eigensolvers can lead respectively to the use
of traditional parallel libraries such as ScaLAPACK or
PARPACK.

In this paper we focus on a specific instance of a plane wave method
which splits the basis functions support domain (Muffin-Tin): in a
spherical symmetric area around each atom labelled by $\alpha$,
$\psi_{\bf G}$ receive contributions by augmented radial functions
$a^{\alpha,{\bf G}}_{\it lm} u^{\alpha}_{\it l} + b^{\alpha,{\bf
    G}}_{\it lm}\dot{u}^{\alpha}_{\it l}$ times the spherical harmonics
$Y_{\it lm}$, while plane waves are supported in the interstitial
space between atoms.
  \begin{eqnarray*}
     \psi_{\bf G}(\kv,\rv) =  & \left\{
	\begin{array}[l]{l}
	\frac{1}{\sqrt{\Omega}}e^{i({\bf k+G})\rv} \qquad \qquad \qquad \qquad \qquad \quad \quad \quad\ -\textrm{Interstitial}\\
	\displaystyle\sum_{\it l,m} \left[a^{\alpha,{\bf G}}_{\it lm}(\kv) u^{\alpha}_{\it l}(r)
	+ b^{\alpha,{\bf G}}_{\it lm}(\kv) \dot{u}^{\alpha}_{\it l}(r) \right] Y_{\it lm}(\hat{\bf r}_{\alpha}) \quad- \textrm{Spheres}. \\
	\end{array}
     \right.
  \end{eqnarray*}
  At each iteration cycle the radial functions $u^{\alpha}_{\it l}(r)$
  are computed anew by solving auxiliary Schr\"odinger equations.
  Moreover a new set of the coefficients $a^{\alpha,{\bf G}}_{\it lm}$
  and $b^{\alpha,{\bf G}}_{\it lm}$ is derived by imposing continuity
  constraints on the surface of the Muffin-Tin spheres.  Consequently
  at each iteration the basis set $\psi_{\bf G}(\kv,\rv)$ changes
  entirely. This discretization of the KS equations -- known as FLAPW
  -- translates in a set of $\mathcal{N}_{\ \kv}$ quite dense GHEVPs
\[
\sum_{\bf G'} (A_\kv)_{\bf GG'}\ c^{\bf G'}_{\kv,i} = \lambda_{\kv,i}
\sum_{\bf G'} (B_\kv)_{\bf GG'}\ c^{\bf G'}_{\kv,i},
\]
each one labeled by a value of the plane wave vector $\kv$. The role
of eigenvectors is played by the $n$-tuple of coefficients $c_{\kv,i}$
expressing the orbital wave functions $\phi_i$ in terms of the basis
wave functions $\psi_{\bf G}$. The entries of each GHEVP matrix are
initialized by evaluating numerically a series of expensive multiple
integrals involving the $\psi_{\bf G}$s. Since we are dealing with
non-linear eigenvalue problems, one resorts to the self-consistent
process described above which now becomes
\\[5mm]
\be
	\begin{array}[l]{ccccc}
 	\color{red}{\fbox{\rule[-0.2cm]{0cm}{0.6cm}\shortstack{{\small Initial input}\\$n_{\rm init}(\rv)$}}} &
	\Longrightarrow & \fbox{\rule[-0.2cm]{0cm}{0.6cm} \shortstack{{\small Compute KS potential}\\$\mathcal{V}_{\rm ef{}f}[\nr]$}} &
	\longrightarrow & \fbox{\rule[-0.2cm]{0cm}{0.6cm} \shortstack{{\small Solve a set of eigenproblems}\\$P^{(\ell)}_{\kv_1} \dots P^{(\ell)}_{\kv_N}$}} \\[4mm]
 	 & & \uparrow {\rm No} & & \downarrow  \\[2mm]
	 \color{blue}{\fbox{\rule[-0.2cm]{0cm}{0.6cm} \shortstack{{\small OUTPUT}\\ {\small Energy, $\dots$}}}} &
	 \shortstack{{\small Yes} \\ $\Longleftarrow$} &  \fbox{\rule[-0.2cm]{0cm}{0.6cm} \shortstack{\small Converged? \\ $|n^{(\ell)} - n^{(\ell-1)}| < \eta$}} &
	 \longleftarrow &  \fbox{\rule[-0.2cm]{0cm}{0.6cm} \shortstack{{\small Compute new density}\\$n(\rv) = \sum_{\kv,i}\, f_{\kv,i} |\phi_{\kv,i}({\bf r})|^2$}}
 	\end{array}
	\nn
\ee
\\[5mm]
There are now $\mathcal{N}_{\ \kv}$ GHEVP for each cycle labeled by $\ell$
\[
P^{(\ell)}_\kv :\quad A^{(\ell)}_\kv\ c^{(\ell)}_{\kv,i} = \lambda^{(\ell)}_{\kv,i}\
B^{(\ell)}_\kv\ c^{(\ell)}_{\kv,i} \qquad (\ell = 1, \dots, N).
\]
All along the sequence the solutions of all $P^{(\ell-1)}_\kv$ are
used to initialize the new eigenproblems $P^{(\ell)}_\kv$. In
particular the eigenvectors $c^{(\ell-1)}_{\kv,i}$ are used to derive
the orbital functions $\phi_{\kv,i}^{(\ell-1)}$ which in turn
contribute to the charge density $n^{(\ell-1)}(\bf r)$. At the next
cycle $n^{(\ell-1)}(\bf r)$ contributes to modify the potential
$\mathcal{V}_{\rm ef{}f}$ which causes the radial functions defining
$\psi^{(\ell)}_{\bf G}$s to change.  This new basis function set
directly determines the initialization of the entries of
$A^{(\ell)}_\kv$ and $B^{(\ell)}_\kv$ and indirectly the new
eigenvectors $c^{(\ell)}_{\kv,i}$. The result is a number
$\mathcal{N}_{\ \kv}$ of sequences of eigenproblems
$\left\{P_\kv\right\}_N$, one sequence for each fixed $\kv$, where the
eigenpairs $(\lambda^{(N)}_{\kv,i},c^{(N)}_{\kv,i})$ converged within
tolerance to the solution of the original non-linear problem.

When solving for an eigenvalue problem the first high level choice is
between direct and iterative eigensolvers. The first are in general
used to solve for a large portion of the eigenspectrum of dense
problems. The latter are instead the typical choice for sparse
eigenproblems or used to solve for just few eigenpairs of dense
ones. In FLAPW the hermitian matrices $A_\kv$ and $B_\kv$ are quite
dense, have size generally not exceeding 30,000, and each
$P^{(\ell)}_\kv$ is solved for a portion of the lower spectrum not
bigger than 20\%. Consequently, when each GHEVP is singled out from
the rest of the sequence, direct solvers are unquestionably the method
of choice. Currently, most of the codes based on FLAPW
methods~\cite{FLEUR,Wien2k, Exciting} use the algorithms BXINV or MRRR
directly out of the ScaLAPACK or ELPA library.  If the use of direct
solvers is the obvious choice when each $P^{(\ell)}_\kv$ is solved in
isolation, the same conclusion may not be drawn when we look at the
entire sequence of $\left\{P_\kv\right\}_N$.


\subsection{Harnessing the correlation}
\label{sec:exploit}

As described in the previous section, the chain of computations that
goes from $P^{(\ell-1)}_\kv$ to $P^{(\ell)}_\kv$ suggests a connection
between eigenvectors of successive eigenproblems. The entries of
$P^{(\ell)}_\kv$ are in fact the result of multiple integrals between
$\psi^{(\ell)}_{\bf G}$ and operators depending on the new charge
density $n^{(\ell-1)}(\bf r)$. All these quantities are
modified by $c^{(\ell-1)}_{\kv,i}$ collectively in such a complex
fashion across each cycle that there is no accessible mathematical
formulation which makes explicit the connection between
$c^{(\ell-1)}_{\kv,i}$ and $c^{(\ell)}_{\kv,i}$. In other words,
according to our definitions, it is clear that each set of problems
$P_\kv^{(\ell)} \quad \ell=1, \dots, N$ is a sequence but there is no
immediate evidence that such sequence is made of correlated
eigenproblems.
\begin{figure}[!htb]
\hspace*{-0.7cm}
\centering
  \subfigure[Evolution of eigenvectors angles.]{
  \includegraphics[scale=0.31]{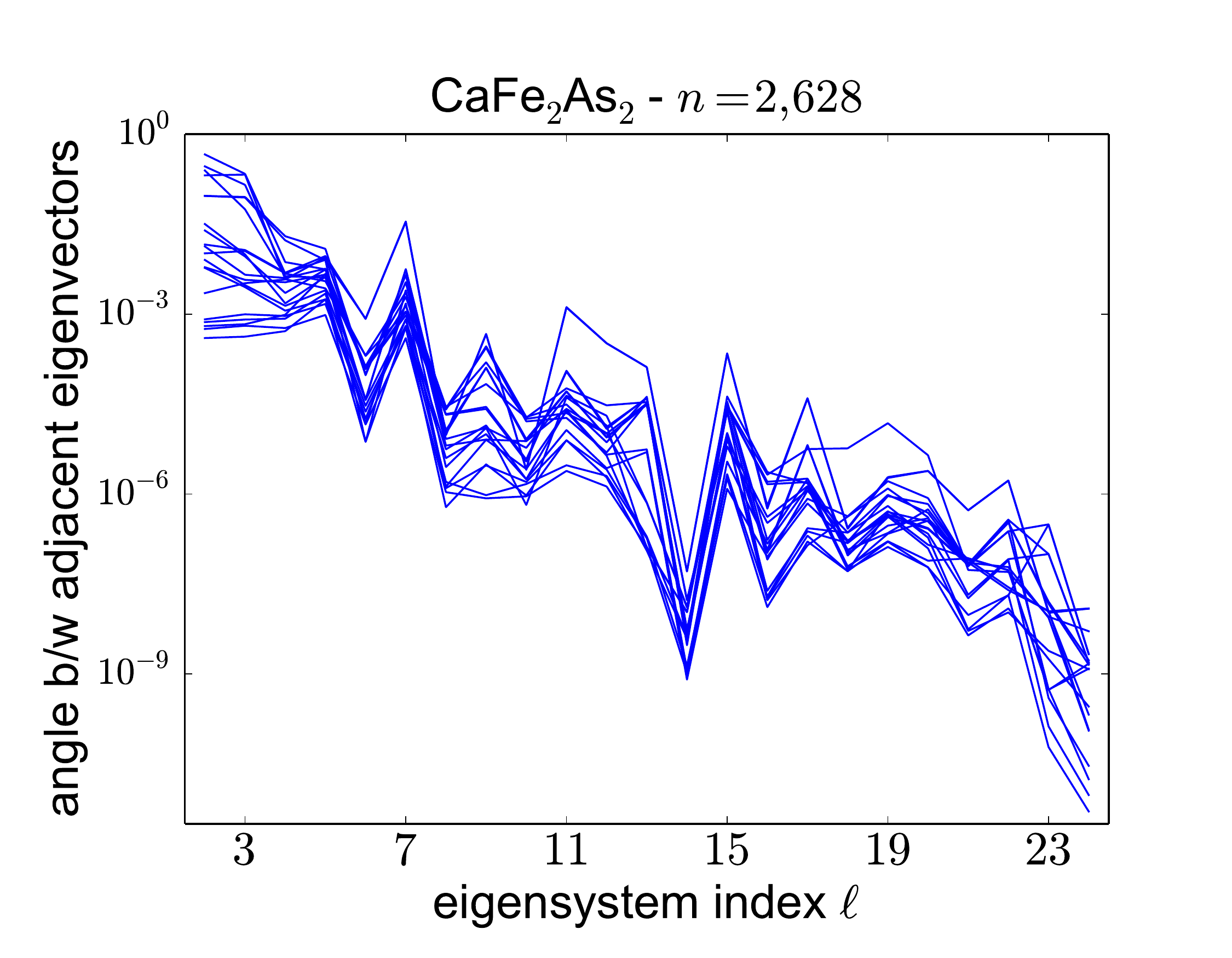}}
  \subfigure[Evolution of eigenspace angle.]{
  \includegraphics[scale=0.31]{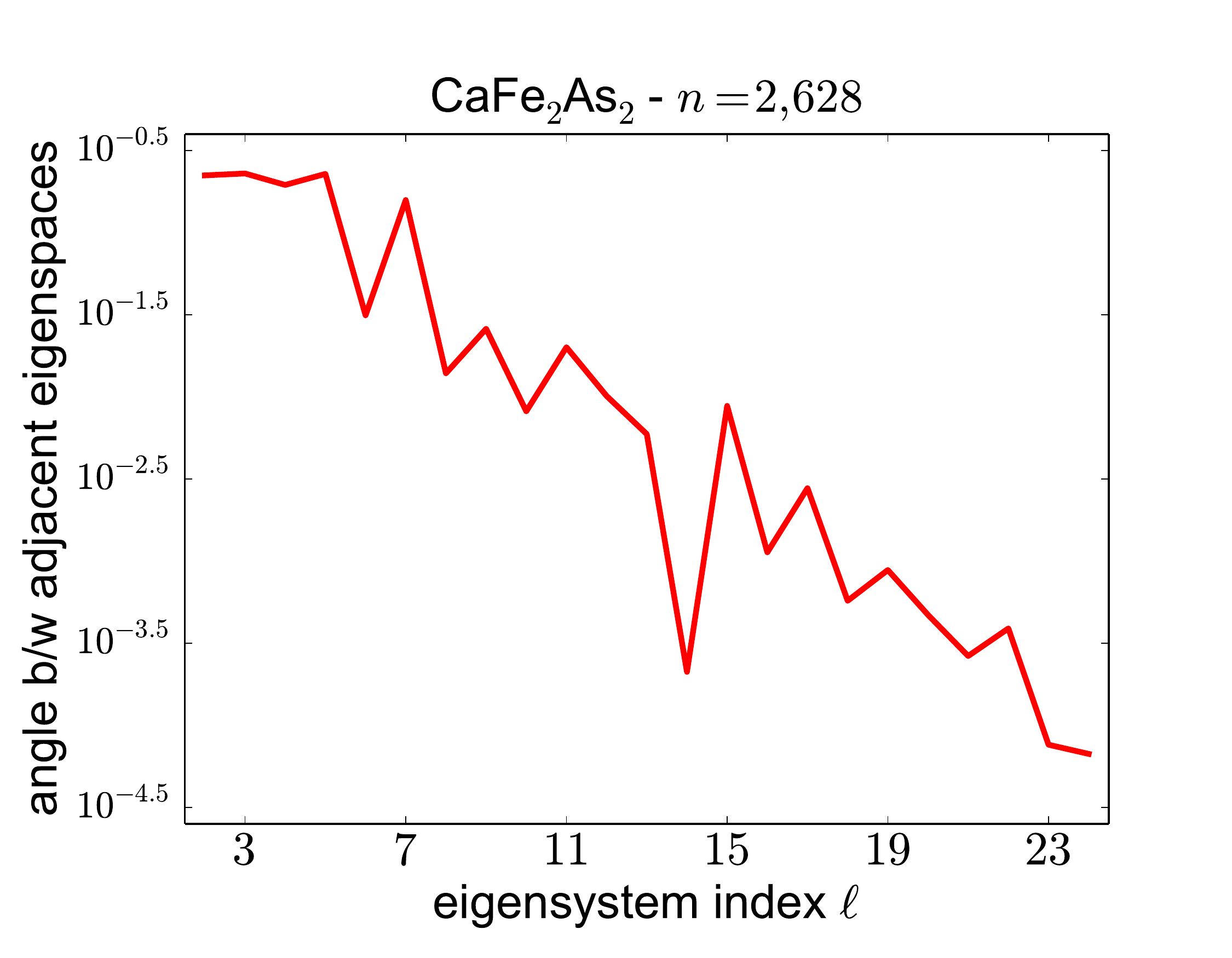}}
\caption{The data in this figure refers to a sequence of eigenproblems
  relative to the same physical system CaFe$_{2}$As$_{2}$. Plot (a)
  represents the angles between eigenvectors corresponding to the
  lowest 20 eigenvalues. Plot (b) shows the angle between subspaces
  spanned by a fraction of the eigenvectors of adjacent
  eigenproblems.}
  \label{fig:angleevol}
\end{figure}

Evidence of the existence of a correlation between the eigenvectors of
successive eigenproblems becomes clear only numerically. As shown
in~\cite{DiNapoli:2012fk} the correlation manifests itself in the
evolution of the angles $\theta^{(\ell)}_{\kv,i} = \left(1 - \langle
  c^{(\ell-1)}_{\kv,i} , c^{(\ell)}_{\kv,i} \rangle\right)$. In
particular the $\theta^{(\ell)}_{\kv,i}$ decrease almost monotonically
as a function of cycle index $\ell$, going from $\sim 10^{-1}$ down to
$\sim 10^{-8}$ towards the end of the sequence (see plot (a) of
Fig.~\ref{fig:angleevol}). Eigenvectors thus become more and more
collinear with increasing $\ell$. Plot (b) of Fig.~\ref{fig:angleevol}
clearly shows that even when measuring the full sought after
eigenspace of adjacent eigenproblems, the entire subspace changes to a
lesser extent as the sequence index grows.

The evolution of the eigenvectors in Fig.~\ref{fig:angleevol} suggests
that they can be ``reused'' as approximate solutions, and inputted to
the eigensolver at the successive cycle. Unfortunately no direct
eigensolver is capable of accepting vectors as approximate
solutions. Therefore if we want to exploit the progressive
collinearity of vectors as the sequence progresses, we are lead to
consider iterative solvers; these solvers by their own nature build
approximate eigenspaces by manipulating approximate eigenvectors. When
inputting to an iterative eigensolver a vector which already
provides a good approximation, it is plausible to expect that the
number of iterations of the solver is reduced and the whole process
speeds-up.

Since we are provided with as many approximate vectors as the
dimension of the sought after eigenspace, the iterative eigensolver of
choice should be able to simultaneously accept multiple vectors as
input. Such a feature is in part provided by the class of block
iterative
eigensolvers~\cite{Cullum:1974fe,Grimes:1994td,Knyazev:2007bn,
Stathopoulos:2007bo,Anonymous:uiOI0oMM,Zhou:2010je,Sadkane:1999uc},
among which one has to select the one that maximize the number of
input vectors and exploit to the maximum extent the approximate guess
they provide~\cite{DiNapoli:2012uy}. Consequently the class of block
solvers is dramatically restricted to subspace iteration
algorithms. While this class of algorithms has the correct properties
it also known to converge at best linearly to an invariant subspace of
the eigenspace~\cite{Golub:2012wt}. By providing the subspace
iteration with a polynomial pre-conditioner the convergence can be
improved substantially~\cite{Rutishauser:1969ub}. Among the many
choices of polynomial accelerators, we singled out the Chebyshev
polynomials for their optimal enhancing of the sought after
subspace. The end product is a Chebyshev Filtered Subspace Iteration
method (ChFSI) which we are going to describe in the next section.


\section{The Chebyshev Filtered Subspace Iteration Method}
\label{sec:ChFSI}

As mentioned in the previous section, when the search for the
solutions to an eigenproblem requires the handling of approximate
eigenvectors, iterative eigensolvers are the obvious choice. Such
solvers attempt to build an invariant subspace of the eigenspace by
repeating in loop-wise fashion a series of steps. At the end of each
loop (or iteration) a subspace, which provides a better guess to the
invariant eigenspace than the previous one, is constructed. Iterative
eigensolvers can be distinguished in two major categories: those for
which the dimension of the constructed subspace changes at the end of
each iteration, and those for which it is maintained fixed. In the
first case the subspace is built out of a single initial vector and
the subspace grows at each iteration until some restart mechanism
resizes it. Krylov- and Davidson-like eigensolvers are all included in
this category. Eigensolvers in the second category have the subspace
spanned by a number of vectors equal (or slightly larger) to the size
of the required spectrum. Typical examples of solvers in this category
are subspace iteration methods.

Block eigensolvers constitute a sort of hybrid group which attempt to
combine the robustness of dimension-changing methods with the
compactness of dimension-invariant ones. A prototypical example of
these methods is a Krylov-like method for which the subspace is
augmented with small blocks of mutually orthogonal vectors. Block
methods are more versatile since they can accept multiple input vectors
to initialize the iterative process. Unfortunately they can also be
more unstable due to the very process of subspace building which can
lead to rank-deficient subspaces. For this reason the size of the
block is never too high and very rarely is larger than 20. Due to
these issues standard block solvers are unsuitable to be used when the
number of input vectors is already of the order of several dozens. The
only alternative is to use subspace iteration methods which, by
definition, are maximal-size block methods.

Besides accepting multiple input vectors, block methods, and
especially subspace iteration, offer a series of additional advantages
when used on dense eigenvalue problems. In the first place they can
solve simultaneously for a portion of the sought after eigenspace. In
particular subspace iteration avoids stalling when dealing with
relatively small clusters of eigenvalues. In the second place, block
methods rely on matrix-matrix rather than matrix-vector multiplication
kernels. When dealing with dense matrices the obvious choice for these
multiplications is the highly optimized {\tt GEMM} routine included
in the BLAS 3 library. Especially in the case of subspace iteration
the use of {\tt GEMM} permits to reach close-to-peak performance of
the processing unit. This is one of the key characteristics that makes
ChFSI a very efficient and easy to scale algorithm.


  \subsection{The sequential algorithm}
  \label{sec:SeqChFSI}

  Subspace iteration is probably one of the earliest iterative
  algorithms to be used as numerical
  eigensolver~\cite{Stewart:1969cq,Jennings:1975tv}. Subspace
  Iteration complemented with a Chebyshev polynomial filter is a well
  known algorithm in the
  literature~\cite{Rutishauser:1969ub,Manteuffel:1977wt,Saad:2011tu}. A
  version of it was recently developed and implemented by Zhou {\sl et
    al.}\footnote{The authors refer to this algorithm as CheFSI not to
    be confused with our ChFSI algorithm.} for a real space
  discretization of DFT~\cite{Zhou:2006ut,Zhou:2006ek} and included in
  the PARSEC code~\cite{Kronik:2006ff}. The ChFSI algorithm here
  presented takes inspiration from this latest implementation and
  includes some additional features: 1) it introduces an internal loop
  that iterates over the polynomial filter and the Rayleigh quotient,
  2) it adds a locking mechanism to the internal loop, and 3) most
  importantly, it optimizes the degree of the polynomial filter so as
  to minimize the number of matrix-vector operations. In addition to
  these supplementary features, ChFSI is specifically tailored for
  sequences of correlated eigenproblems. In fact the algorithm can be
  used to solve an isolated eigenvalue problem but thrives when it is
  inputted with knowledge extracted from the sequence.  The ChFSI
  pseudocode is illustrated in {\bf Algorithm~\ref{alg:ChFSI}}. Notice
  that the initial input is not the initial $P^{(\ell)}$ but its
  reduction to standard form $H^{(\ell)} =
  L^{-1}A^{(\ell)}L^{-\textsc{T}}$ where $B^{(\ell)}=LL^{\textsc{T}}$,
  and $Y^{(\ell-1)}$ are the eigenvectors of $H^{(\ell-1)}$.

It is a known fact that any implementation based on subspace iteration
converges linearly at best. By using a polynomial filter on the
initial block of inputted vectors the method experiences a high rate of
acceleration. In order to implement an efficient filter, ChFSI uses
few Lanczos iterations ({\tt line~\ref{lst:line:lanczos}}) so as to
estimate the upper limit of the eigenproblem
spectrum~\cite{Zhou:2011ic}. This estimate constitutes the upper bound
of the interval whose eigenvectors are suppressed by the filter. The
lower bound is given by the inputted
$\lambda^{(\ell-1)}_{\textsc{NEV}+1}$ of the previous iteration cycle. These
two values are necessary for the correct usage of the filter based on
Chebyshev polynomials~\cite{Saad:2011tu}.

\newlength{\clen}
\settowidth{\clen}{{\sc rayleigh-ritz} (Start)}
\newlength{\ylen}
\settowidth{\ylen}{Y}

\begin{algorithm}[h!t]
  \caption{Chebyshev Filtered Subspace Iteration}
  \label{alg:ChFSI}
  \begin{algorithmic}[1]
    \Require Matrix $H^{(\ell)}$, approximate eigenvectors
    $Y^{(\ell-1)}$ and eigenvalues $\lambda^{(\ell-1)}_1$ and
    $\lambda^{(\ell-1)}_{\textsc{nev}+1}$.  Starting degree {\sc deg}
    for the filter.  \Ensure Wanted eigenpairs
    $\big(\Lambda^{(\ell)},Y^{(\ell)}\big)$.
  \item[] \State Set $\big(\Lambda^{(\ell)},Y^{(\ell)}\big)$ to be the
    empty array/matrix.  \State Set the degrees $(m_1, \dots,
    m_{\textsc{nev}}) = (\textsc{deg}, \dots, \textsc{deg})=:m$ for
    the filter.  \State Estimate the largest
    eigenvalue. \label{lst:line:lanczos} \Comment{\parbox{\clen}{\sc
        lanczos}} \Repeat \State Filter the vectors
    $Y^{(\ell-1)}=C_{m}\big(Y^{(\ell-1)}\big)$. \label{lst:line:cheby}
    \Comment{\parbox{\clen}{\sc chebyshev filter}} \State
    Orthonormalize
    $\big[Y^{(\ell)}\;Y^{(\ell-1)}\big]$. \label{lst:line:QR}
    \Comment{\parbox{\clen}{\sc qr}} \State Compute Rayleigh quotient
    $G=Y^{(\ell-1)\dagger}H^{(\ell)}Y^{(\ell-1)}$. \label{lst:line:rrstarts}
    \Comment{\parbox{\clen}{{\sc rayleigh-ritz} (Start)}} \State Solve
    the reduced problem $GW=W\Lambda^{(\ell-1)}$.  \State Compute
    $Y^{(\ell-1)} = Y^{(\ell-1)}W.$ \label{lst:line:rrends}
    \Comment{\parbox{\clen}{{\sc rayleigh-ritz} (End)}} \State Lock
    converged eigenpairs into $\big(\Lambda^{(\ell)},Y^{(\ell)}\big)$.
    \Comment{\parbox{\clen}{\sc locking}} \State Compute the degrees
    $(m_1, \dots, m_{k})=m$ for the filter.\label{lst:line:opt}
    \Comment{\parbox{\clen}{\sc optimization}} \Until{ converged
      $\geq$ \textsc{nev}}
  \end{algorithmic}
\end{algorithm}

After the Chebyshev filter step ({\tt line~\ref{lst:line:cheby}}) the
block of vectors spanning the invariant subspace could easily become
linearly dependent. In order to avoid such an occurrence each
iteration is usually complemented with some orthogonalization
procedure which, in ChFSI, is established with a Householder reflectors based
QR factorization ({\tt line~\ref{lst:line:QR}}). The $Q$ vectors of the
factorization are then used to compute the Rayleigh-Ritz quotient of
the matrix $H^{(\ell)}$ ({\tt line~\ref{lst:line:rrstarts}}). Such a
quotient represents a projection of the eigenproblem onto a subspace
approximating the sought after eigenspace. The resulting reduced
eigenproblem is then diagonalized and the computed eigenvectors are
projected back to the larger problem ({\tt
  line~\ref{lst:line:rrends}}). At the end of the Rayleigh-Ritz step
eigenvector residuals are computed, converged eigenpairs are locked
and the non-converged vectors are set to be filtered again. For each
non-converged vector, an optimized degree of the polynomial filter is
computed using its corresponding residual and approximate eigenvalue
({\tt line~\ref{lst:line:opt}}).

\begin{figure}[!htb]
\hspace*{-0.7cm}
\centering
\subfigure[EleChFSI steps timing profile.]{
\includegraphics[scale=0.31]{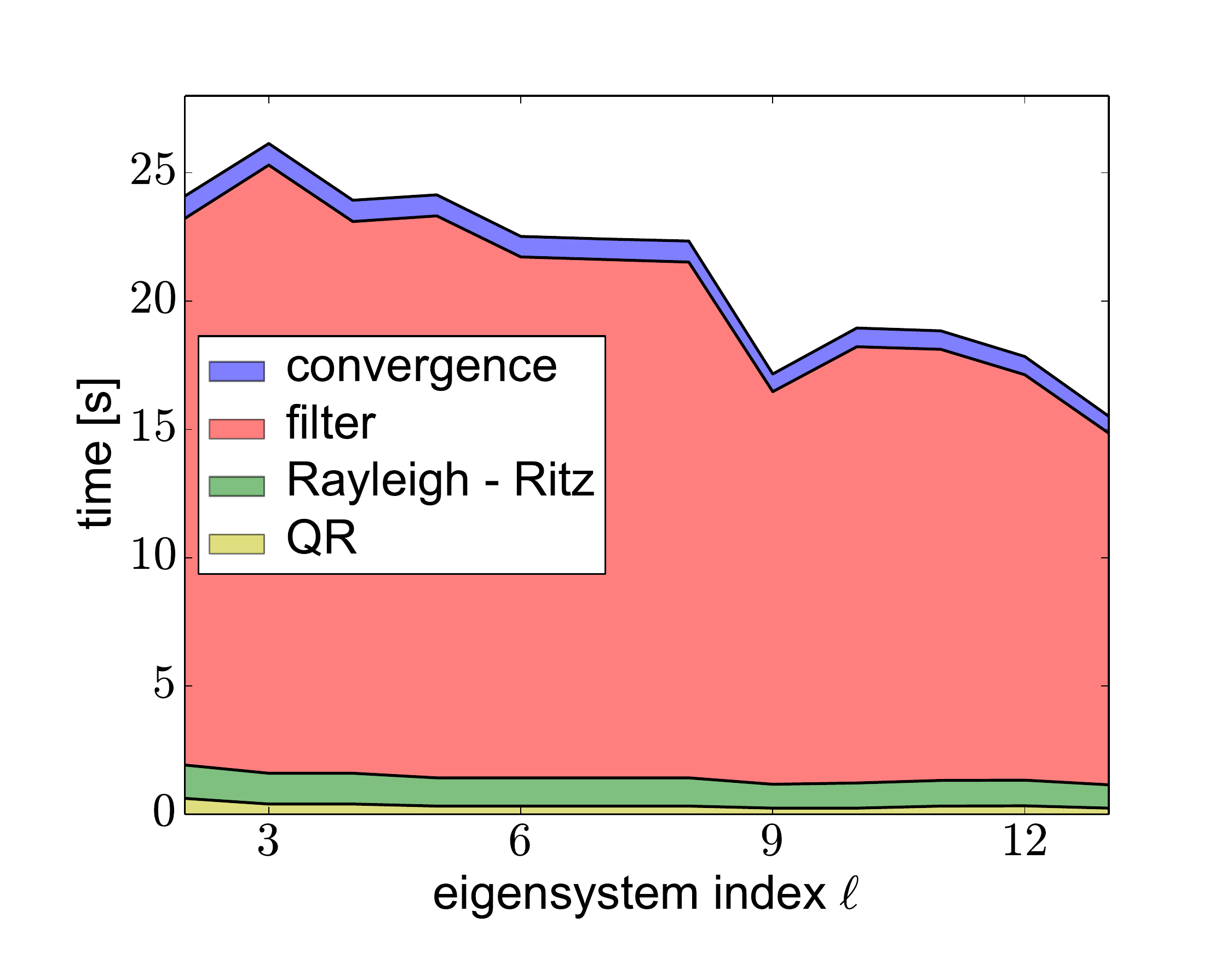}}
\subfigure[Random vs Approximate vectors.]{
\includegraphics[scale=0.31]{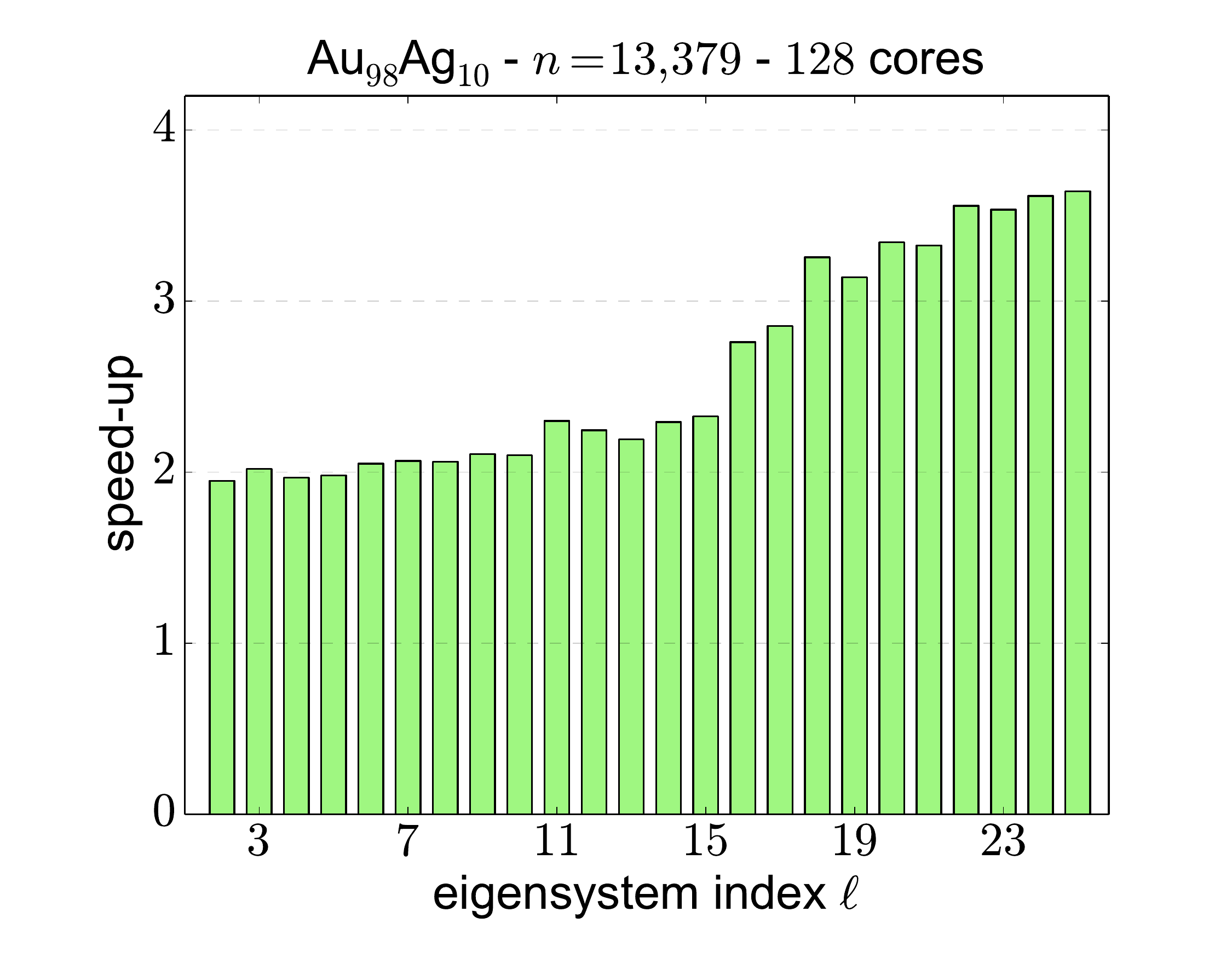}
}
\caption{EleChFSI behaviour as a function of the iteration
  (eigenproblem) index $\ell$. In plot (a) it is depicted the timing
  profile for each of the steps. Plot (b) shows the speed-up of
  EleChFSI when inputted approximate solutions as opposed to random
  vectors.}
\label{fig:timeprof}
\end{figure}

As for any iterative eigensolver, it is not feasible to predict how
many loops ChFSI requires to find all desired eigenpairs. Consequently
it is not possible to a priori establish the full computational cost
of the eigensolver. In its stead it is feasible to evaluate the cost
of each step within a single loop. Let's express with $m$ the degree
of the polynomial, $n$ the size of the eigenproblem and $k$ the number
of filtered vectors. Then the complexity of the {\sc chebyshev filter}
is $\odg{mn^2k}$, the {\sc qr} factorization accounts for
$\odg{nk^2}$, the whole {\sc rayleigh-ritz} procedure amounts to
$\odg{n^2k+nk^2+k^3}$, and the residuals check in the {\sc locking}
step is $\odg{n^2k}$~\footnote{The {\sc lanczos} and {\sc
    optimization} steps are at most $\odg{n^2}$ and $\odg{k}$
  respectively and are consequently negligible.}.  Since $m \gg 1$,
the filter makes up for the highest fraction of computing
time. Moreover more loops, and so more filtering steps, will be needed
to solve for problems at the beginning of the sequence, making the
filter complexity even more dominant there. In plot (a) of
Fig.~\ref{fig:timeprof} we show how this heuristic analysis is
confirmed by numerical measurements\footnote{Data in the plot refers
  to the parallelized version of ChFSI which is discussed in a later
  section.}.

Many of the most expensive steps of the algorithms can leverage the
{\tt GEMM} kernel of the BLAS 3 library. In the Rayleigh-Ritz, {\tt
  GEMM} can be invoked for both the computation of the quotient and
the back-transformation of the Ritz vectors. The computation of the
vectors residuals can also profit from BLAS 3 calls. As we will see in
the next section, the filter is the part of the algorithm which
benefits by far the most from the usage of this dense linear algebra
kernel.

\newlength{\fleno}
\settowidth{\fleno}{$\sigma_1$}
\newlength{\flenp}
\settowidth{\flenp}{$Y_{i+1,1\mathtt{:s}-1}$}
\newlength{\flenq}
\settowidth{\flenq}{$\sigma_{i+1}$}

\begin{algorithm}[h!t]
  \caption{Chebyshev Filter}
  \label{alg:filter}
  \begin{algorithmic}[1]
    \Require Matrix $H\in\mathbb{C}^{n\times n}$, vectors $Y_0\in\mathbb{C}^{n\times k}$
    sorted according to the ascending degree specification $m=(m_1, \dots, m_k)\in\mathbb{N}^{k}$
    and parameters $\lambda_1, c, e\in\mathbb{R}$.
    \Ensure Filtered vectors $Y_{m}$, where each vector $Y_{m, j}$ is
    filtered with a Chebyshev polynomial of degree $m_j$.
  \item[]
    \State $\parbox{\fleno}{H} \leftarrow \left(H-cI_N\right)/e$
    \State $\parbox{\fleno}{$\sigma_1$} \leftarrow e/\left(\lambda_1-c\right)$
    \State $\parbox{\fleno}{$Y_1$} \leftarrow \sigma_1HY_{0}$
    \State $\parbox{\fleno}{$\mathtt{s}$} \leftarrow \argmin_{j=1,\ldots,k}m_j\neq 1$
    \For{$i=1,\ldots,m_k-1$}
    \State $\parbox{\flenq}{$\sigma_{i+1}$} \leftarrow 1/\left(2/\sigma_1-\sigma_i\right)$
    \State $\parbox{\flenp}{$Y_{i+1,1\mathtt{:s}-1}$} \leftarrow Y_{i,1\mathtt{:s}-1}$
    \State $\parbox{\flenp}{$Y_{i+1,\mathtt{s:}n}$} \leftarrow 2\sigma_{i+1}HY_{i,\mathtt{s:}n}-\sigma_{i+1}\sigma_{i}Y_{i,\mathtt{s:}n}$ \label{lst:line:3term}
    \State $\mathtt{s}\leftarrow \argmin_{j=\mathtt{s},\ldots,k}m_j\neq i+1$
    \EndFor
  \end{algorithmic}
\end{algorithm}


  \subsubsection{The Chebyshev filter}
  \label{sec:filter}

Since the Chebyshev filter is heavily based on the use of the
homonymous polynomials, let us recall briefly their definition and
properties.
\begin{definition}
  The Chebyshev Polynomials $C_m(t)$ of degree $m$ can be defined on
  the entire real axis as
  \begin{equation}
   C_m(t)=\cosh\left(m\cosh^{-1}(t)\right), \quad t\in\mathbb{R}.
   \label{eq:polydef}
  \end{equation}
  When the domain of definition is restricted on the interval
  $\left[-1,1\right]$ their expression simplifies to
  \begin{equation*}
   C_m(t)=\cos\left(m\cos^{-1}(t)\right), \quad \left| t\right| \leq 1.
  \end{equation*}
\end{definition}
Despite the definition hides their polynomial nature, it can be easily
shown that $C_m(t)$ oscillates between the values $-1$ and $1$ in the
interval $t\in [-1, 1]$ and diverge monotonically taking values
\[
C_m(t) = \frac{|\rho|^m+|\rho|^{-m}}{2}, \qquad |\rho| = \max_{|t|
  >1} \left|t \pm \sqrt{t^2-1}\right|.
\]
Additionally, the combination $p_m(t) = \left.
\frac{C_m(t)}{C_m(\gamma)}\right|_{|\gamma|\geq 1}$ is shown to
satisfy an extremum theorem
(See Th.~4.8 in \cite{Saad:2011tu}). The theorem together with the
asymptotical behaviour of $C_m(t)$ determine the convergence properties
of $p_m(A)$ when it applies to a generic vector $y$. Assume $A
\in\mathbb{C}^{n\times n}$ is Hermitian, $(\lambda_i, w_i)$ are its
eigenpairs in ascending order and that the interval $[-1, 1]$ is
mapped to an interval $[\alpha, \beta] \ni \{\lambda_2, \dots,
\lambda_n\}$, it is then straightforward to show that
\[
p_m(A) y = \sum_{i=1}^n s_i p_m(\lambda_i) w_i \approx s_1w_1 + \sum_{i=2}^ns_i \frac{1}{\left|\rho_1\right|^m}w_i,
\]
which converges to $w_1$ with ratio
\begin{equation}
  \tau(\lambda_1)
  = |\rho_1|^{-1}
  = \min_{\lambda_1\notin [\alpha,\beta]}\left\lbrace
    \left|\lambda_{1,c,e} +\sqrt{\lambda_{1,c,e}^2-1}\right|,
    \left|\lambda_{1,c,e} -\sqrt{\lambda_{1,c,e}^2-1}\right|
  \right\rbrace,
  \label{eq:ratio}
\end{equation}
where $c=\frac{\alpha+\beta}{2}$ and $e=\frac{\beta-\alpha}{2}$ are
respectively the center and the half-width of the interval $[\alpha,
\beta]$ and $\lambda_{1,c,e}=\frac{\lambda_1-c}{e}$. The further is
the eigenvalue $\lambda_1$ from the interval, the smaller is
$\tau(\lambda_1)$ and so the faster is the convergence to $w_1$.

{\bf Algorithm~\ref{alg:filter}} generalizes the mechanism just
described above to the case of an arbitrary number $k$ of vectors $Y$
approximating the eigenvectors corresponding to the values
$\{\lambda_1, \dots, \lambda_k\} \notin [\alpha, \beta]$. The values
for $\beta$ and $\alpha$ are respectively estimated by the {\sc
  lanczos} step of {\bf Algorithm~\ref{alg:ChFSI}} and by
$\lambda^{(\ell-1)}_{\textsc{nev}+1}$\footnote{This is only true for
  the first loop of ChFSI. After that it is given by the approximate
  value $\hat{\lambda}^{(\ell)}_{\textsc{nev}+1}$ given by the
  previous loop.}, while $\lambda_1^{(\ell - 1)}$ gives an estimate
for $\gamma$. The actual polynomial $p_m(t)$ is not computed
explicitly. What is calculated is its action on the vectors $Y$ by
exploiting the $3$-terms recurrence relation which can be also used to
define Chebyshev polynomials
\begin{equation}
\label{eq:3term}
C_{m+1}(Y) = 2HC_m(Y) - C_{m-1}(Y),
\qquad C_m(Y) \equiv C_m(H)\cdot Y.
\end{equation}
Eq.~(\ref{eq:3term}) generalizes easily to $p_m(Y)$ as shown in {\tt
  line \ref{lst:line:3term}} of {\bf Algorithm~\ref{alg:filter}}.

The choice of polynomial degree deserves a discussion on its
own. Although the ratio of convergence for each vector depends on the
corresponding eigenvalues as in \eqref{eq:ratio}, we can calculate a
reasonably accurate estimate of it. This claim is based on two
observations. First we start already with an approximate eigenspectrum
which is rapidly refined at each additional loop. Second, it is known
that the eigenvalues convergence goes as the square of the convergence
of their corresponding eigenvectors. These two facts justifies the use
of the values $\tau(\hat{\lambda}_i) = |\hat{\rho}_i|^{-1}$, computed
with approximate eigenvalues $\hat{\lambda}_i$ after the first loop,
as already good estimates for the exact ratios $|\rho_i|^{-1}$.  As
shown in~\cite{BerDin-toappear}, estimates for the ratios can be used
in modeling the behaviour of the filtered vectors residuals
\begin{equation}
	\begin{array}{rll}
          \textrm{Res}(Y_{m,1}) \doteq &  {\|H Y_{m,1} - \hat{\lambda}_1 Y_{m,1}\| \over \|Y_{m,1}\|} \lesssim & \frac{\left| \textrm{const.}\right|}{\left|\hat{\rho}_1\right|^m} \\[6mm]
          \textrm{Res}(Y_{m,i}) \doteq & {\|H Y_{m,i} - \hat{\lambda}_i Y_{m,i}\| \over \|Y_{m,i}\|} \lesssim &
          \frac{\left| \textrm{const.} \right|}{\left|\hat{\rho}_i\right|^{m}} +
          \left| \textrm{const.} \right| \frac{\left|\hat{\rho}_1\right|^{m}} {\left|\hat{\rho}_i\right|^{m}}\, \cdot \varepsilon, \qquad 2 \leq i \leq k,
	\end{array}
\label{eq:resl}
\end{equation}
where with $\varepsilon$ we refer to a value which could be as small
as the machine-epsilon.

Eq.~(\ref{eq:resl}) portraits two regimes for the eigenpair residuals
with $i \geq 2$: a converging regime for small to moderate values of
$m$, and a diverging regime for moderate to large values of $m$. In
the converging regime the first term dominates by monotonically
decreasing the residuals to lower values. The diverging regimes kicks
in when the ratio
$\frac{\left|\hat{\rho}_1\right|^{m}}{\left|\hat{\rho}_i\right|^{m}}
\gg 1$ and, counterbalancing the machine precision, contributes to the
residual with a value comparable to the first term. At this point the
second term becomes dominant and increases monotonically the value of
the residuals.

When we are in the converging regime and the initial residual of each
input vector $Y_i$ is known, Eq.~(\ref{eq:resl}) can be used to
compute an upper bound for the degree of the polynomial $m_i$ needed
to compute eigenpairs having residual lower than a required tolerance
$\textrm{Res}(Y_{m_i,i}) \leq \textsc{tol}$. Reasonable values for
$|\hat{\rho}_i|$ can be computed at the end of the first loop with a
fairly low polynomial degree $m_0 < 10$.
Our experience showed that even with such a low
degree the computed ratios were fairly accurate. Then by imposing
\[
\textsc{TOL} \geq \textrm{Res}(Y_{(m_i+m_0),i}) \approx \textrm{Res}(Y_{m_0,i}) 
\frac{\left|\hat{\rho}_i\right|^{m_0}}{\left|\hat{\rho}_i\right|^{m_i+m_0}}
\]
we arrive at the following inequality for the polynomial degree which
is used to filter the non-converged vectors at the next loop
\begin{equation}
  m_i \gsim \frac{\ln\left|\frac{\textrm{Res}(Y_{m_0,i})}{\textsc{tol}}\right|}{ \ln\|\hat{\rho}_i\|}.
\end{equation}
Clearly to be effective the computed $m_i$s have to be within the limits
of the converging regime which, 
in our experience, appears always to be the case for values of $m_i$
lower than 40.
  

As we have seen at the beginning of this section, the complexity of
the filter is linearly proportional to the degree of the polynomial
used. Consequently minimizing such a degree boils down to lower the
number of matrix-vector operations to a minimum.  We refer to such a
minimization procedure as \emph{Single Optimization}. Another layer of
optimization is accessible when the whole eigenproblem sequence is
taken into consideration. We refer to such second layer as
\emph{Multiple Optimization}, and direct the reader to
\cite{BerDin-toappear} for further analysis and discussions on
both optimization schemes.

Practically all operations in the filter are carried on using the BLAS
3 subroutine {\tt GEMM}. Since {\tt GEMM} is universally recognized as
the most performant routine available (in some cases reaching up to
95\% of theoretical peak performance), the filter is not only the most
computationally intensive part of the algorithm ((see plot (a) in
Fig.~\ref{fig:timeprof}) but also the most efficient and potentially a
very scalable one.


  \subsection{Parallelizing ChFSI}
  \label{sec:parchfsi}

  The algorithm can be efficiently implemented for both shared and
  distributed memory architectures. Here, we report only the
  latter. The interested reader can find details on an initial
  \footnote{Lacks the degree optimization.}  OpenMP based shared
  memory implementation in ~\cite{berl12, DiNapoli:2012uy}.  The
  parallel distributed memory version is implemented, using the
  Message Passing Interface (MPI), on top of Elemental, a (relatively
  new) distributed memory dense linear algebra
  library~\cite{Poulson:2013fs}.  The parallel algorithm was
  consequently called EleChFSI.  In the following we first comment on
  the design of Elemental, and then address its usage for the
  implementation of EleChFSI.

  {\bf Elemental ---} The library is implemented using C++ and
  provides routines for both sequential and parallel dense linear
  algebra relying heavily on BLAS and LAPACK. It exploits the
  object-oriented nature of C++ providing the \texttt{Matrix} class,
  which internally ``hides'' the details about the matrix datatype
  (functions and operators are overloaded), size, leading dimension
  and so on.  The net effect is to lift the user from the burden of
  passing all those attributes to internal routines as it is customary
  in BLAS and LAPACK libraries. In fact, the provided \texttt{blas}
  and \texttt{lapack} namespaces represent an easy-to-use,
  datatype-independent wrapper over a collection of the most used BLAS
  and LAPACK routines (others can be called directly on the buffer
  where the matrix is stored, if needed).  

  The parallel analogue of the \texttt{Matrix} class is the
  \texttt{DistMatrix} class which additionally internally handles
  distribution details such as alignments and grid information. The
  essential difference in the design of Elemental compared to
  ScaLAPACK/PLAPACK is that Elemental performs all computations using
  element-wise (``elemental'' or ``torus-wrap'') matrix
  distributions. Among the many different distributions supported, we
  used the two dimensional cyclic element-wise one.  Here, two
  dimensional means that the MPI processes involved in the computation
  are logically viewed as a two-dimensional $r\times c$ process grid
  where $p=r\cdot c$ is the total number of processors. Each processor
  is labeled with two indices.  A matrix
  $A=[a_{ij}]\in\mathbb{F}^{n\times m}$ is distributed over the grid
  in such a way that the process $(s,t)$ owns the matrix
\[
A_{s,t}=
\begin{pmatrix}
  a_{\gamma,\delta} & a_{\gamma,\delta+c} & \hdots\\
  a_{\gamma+r,\delta} & a_{\gamma+r,\delta+c} & \hdots\\
  \vdots & \vdots &
\end{pmatrix},
\]
where $\gamma \equiv \left(s+\sigma_r\right)\pmod{r}$ and $\delta
\equiv \left(t+\sigma_c\right)\pmod{c}$, and $\sigma_r$
and $\sigma_c$ are arbitrarily chosen alignment parameters.
We will hereafter refer to this distribution as the default one.

The library provides \texttt{View} and \texttt{Partition}
mechanisms to allow the user to work with sub-matrices. The
routines work purely on a pointer basis and automatically handle the
distribution details eliminating the need of allocating additional
memory space where to copy the data.
The communication for the computations is performed almost entirely in
terms of collective communication within rows and columns of the
process grid. Elemental is a constantly growing, modern alternative
to ScaLAPACK/PLAPACK which not rarely outperforms the two precursors.

\begin{figure}[!htb]
\hspace*{-0.9cm}
\centering
  \includegraphics[scale=0.34]{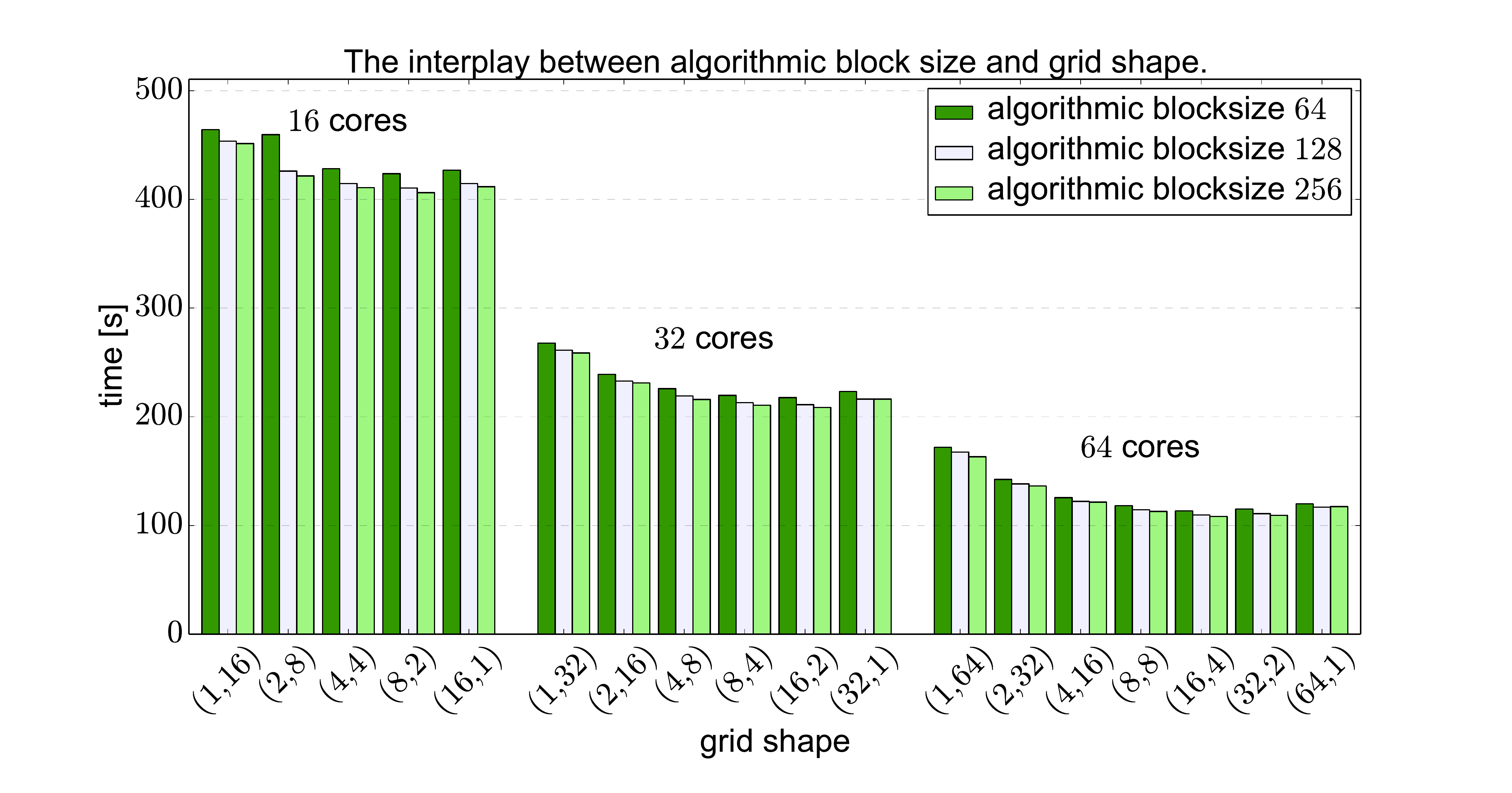}
  \caption{The data in this plot refer to a GHEVP of $\ell=20$, size
    $n=13,379$, and number of sought after eigenpairs $\textsc{nev} =
    972$, corresponding to the physical system Au$_{98}$Ag$_{10}$. The
    eigenproblem was repeatedly solved with EleChFSI using $16$, $32$, and
    $64$ cores, all possible grid shapes $(r,c)$ and three distinct
    algorithmic block sizes.}
  \label{fig:histogram}
\end{figure}

{\bf EleChFSI ---} The Hamiltonian, as well as the approximate
eigenvectors are distributed over the two dimensional grid in the
default manner. For a fixed number $p>1$ of processors there are
several possible choices for $r$ and $c$ forming different grid shapes
$(r,c)$.  The grid shape can have a significant impact on the overall
performance, and careful experiments need to be undertaken in order to
determine the best choice.  Another parameter which affects
performance is the algorithmic block size, which is correlated to the
square root of the L2 cache~\cite{Goto:2008im}.  In practice, the size
of the algorithmic block not only depends on the algorithm itself, but
it is also affected by the architecture and the used compilers.  In
Figure \ref{fig:histogram} we show the performance of EleChFSI with
respect to different grid shapes and different algorithmic block
sizes.  Since we are solving for a small fraction of the
eigenspectrum, the matrix of vectors $Y^{(\ell)}$ is tall and skinny
and we expect a narrow rectangular grid shape to do a better job than
a square and wider one. This is confirmed by the tests.  Furthermore,
the Figure shows that for EleChFSI a block size of 256 is always
recommended independently of the number of cores or grid shape. This
effect is imputable to the large number of matrix multiplications
carried on by the filter.  

The core of EleChFSI is {\bf Algorithm~\ref{alg:filter}}. In the very
first iteration $m_1=m_2=\ldots=m_{\textsc{nev}}$ and the filter
reduces to the ``conventional'' one, where each vector is filtered
with the same degree $\textsc{deg} \equiv m_0$.  Due to the
introduction of \emph{Single Optimization}, in subsequent iterations
these degrees may differ and, in order to exploit {\tt GEMM}
maximally, the vectors are sorted according to their increasing
polynomial degree. The sorting allows us to invoke {\tt GEMM} $m_1$
times on the whole block of vectors $Y_0$, then additional $m_2-m_1$
times on the right block of $Y_{m_1}$, excluding the vectors with
degree $m_1$ which have been already filtered. This process continues
until we filter only the last vector $m_k$ times, as it is illustrated
in {\bf Algorithm~\ref{alg:filter}}.  

The reduced eigenproblem in the Rayleigh-Ritz step is solved using a
parallel implementation of the MRRR eigensolver --
EleMRRR~\cite{Petschow:2013vc} -- which is also integrated in
Elemental. Since the evaluation of the degrees $m_i$ is based on a
heuristic model and some approximate eigenvectors can be better
approximations than others, it is not always the case that a smaller
eigenpair will converge before a bigger one. Therefore, we bring the
converged eigenvectors to the leftmost part of the matrix $Y$ and
leave the non-converged in the remaining rightmost part. A
\texttt{View} of $Y$ then makes sure we work only with the part we are
interested in.


  \section{Numerical tests and discussion}
  \label{sec:numtest}

  The set of numerical tests presented in this section were performed
  on eigenproblems extracted from the simulation of four distinct
  physical systems. The simulations were carried out using the FLEUR
  code~\cite{FLEUR} implementing the FLAPW method. Each of the
  physical systems generated four different eigenproblem sequences of
  distinct length $N$ and size $n$. The data of the sequences of
  eigenproblems is summarized in Table \ref{tab:sim}. 

  By including both conductors and insulators, the physical systems
  represent a heterogeneous sample with distinct physical
  properties. For example the two TiO$_2$ compounds are built using
  respectively $3\times3\times3$ supercell with 161 atoms, and a
  $4\times4\times4$ supercell with 383 atoms, with a single oxygen
  defect. The valence electrons are formed by the Oxygen $2s$ and $2p$
  electrons and the Titanium $3p$, $4s$ and $3d$ electrons resulting
  in a total valence charge of $1,182$ electrons for the small and
  $2,816$ electrons for the large system. The functions basis set
  contained $12,293$ augmented plane waves (APW) and additionally
  $162$ local orbitals (LO), which were increased to $29,144$ APW +
  $384$ LO for the larger system.
  
  \begin{table}[ht]
  \caption{ Simulation data}
  \centering
  \begin{tabular}{c c c c }
  \hline \hline \\[-3mm]%
  Material & $N$ & $n$ \\ [0.5ex]
  \hline \\[-3mm]%
  Na$_{15}$Cl$_{14}$Li & 13 & 9,273 \\
  TiO$_2$ & 30 & 12,455 \\
  Au$_{98}$Ag$_{10}$ & 25 & 13,379 \\
  TiO$_2$ & 18 & 29,528 \\
  \hline%
  \end{tabular}\\
  \label{tab:sim}
  \end{table}

  All our numerical tests were performed on JUROPA, a large general
  purpose cluster where each node is equipped with 2 Intel Xeon X5570
  (Nehalem-EP) quad-core processors at 2.93 GHz and 24 GB memory
  (DDR3, 1066 MHz).  The nodes are connected by an Infiniband QDR
  network with a Fat-tree topology. The tested routines were compiled
  using the GCC compilers (ver. 4.8.2) and linked to the ParTec's
  ParaStation MPI library (ver. 5.0.28-1). The Elemental library
  (ver. 0.84-dev) was used in conjunction with Intel's MKL BLAS and
  LAPACK (ver 11.0). 

  Statistically significant measurements were obtained by running each
  numerical test multiple times and taking the average of the recorded
  CPU times. In order to probe EleChFSI at different level of
  accuracy, eigenproblems were solved setting the minimum tolerance
  \textsc{tol} for the eigenpair residuals to either $10^{-08}$ or
  $10^{-10}$. As already mentioned in Sec.~\ref{sec:parchfsi} the
  choice of processors grid shape and algorithmic block size is
  crucial to obtain the best performance from the Elemental
  framework. Figure \ref{fig:histogram} illustrate that the optimal
  choices of grid shape and algorithmic block size -- for $p = 2^q$
  processor on the JUROPA architecture -- are respectively $(2^{q-2},
  4)$ and 256. All the tests were carried out using exclusively these
  values.

\subsection{Parallel performance and scalability} 

In this subsection we momentarily abstract from the eigenproblem
sequence and test the scalability and parallel efficiency of the
eigensolver for problems at fixed $\ell$. Tests were conducted on
eigenproblems extracted from sequences arising from the first three
chemically different physical systems listed in Table \ref{tab:sim}.

\begin{figure}[!htb]
\hspace*{-0.7cm}
\centering
\subfigure[Strong scalability.]{
\includegraphics[scale=0.31]{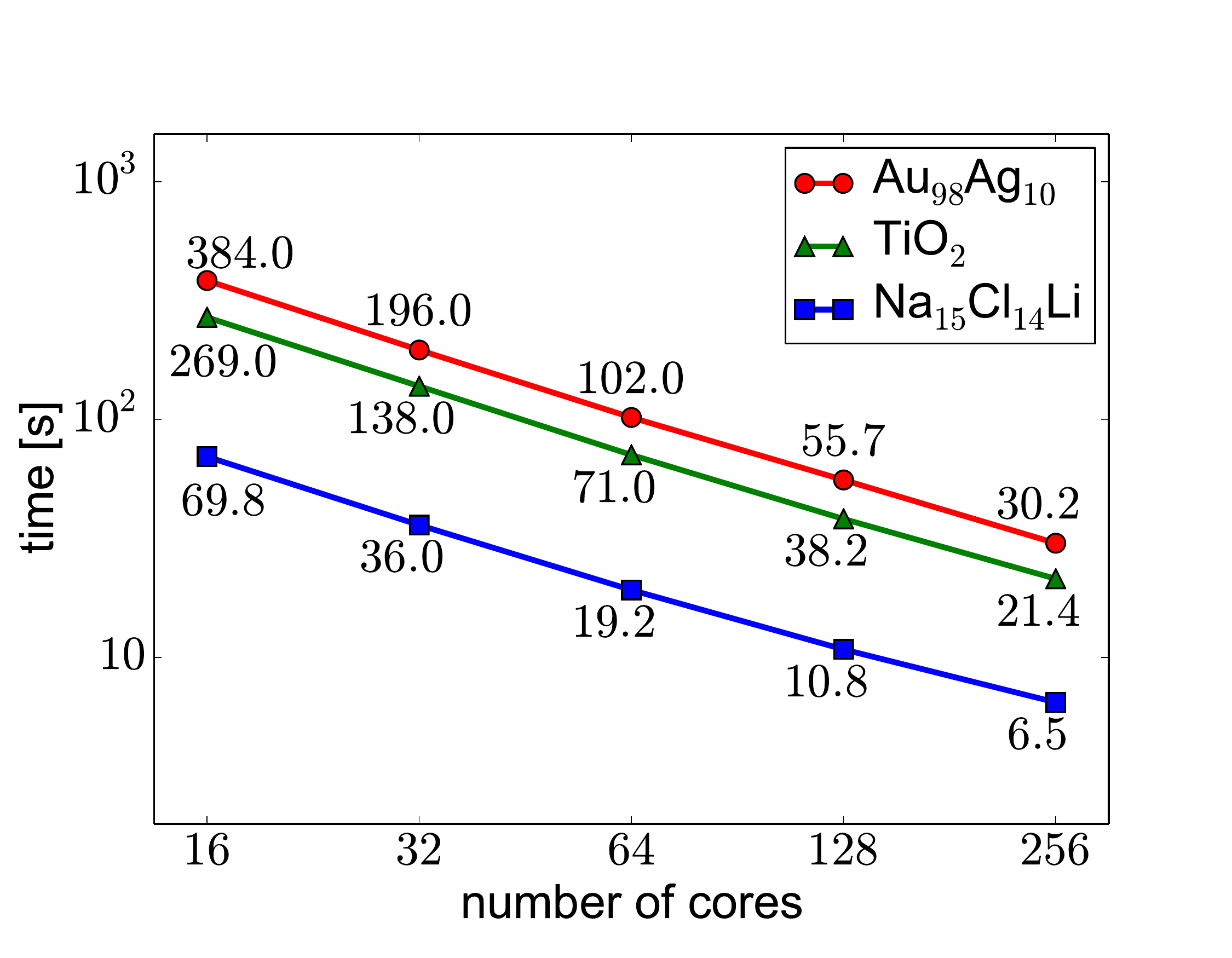}}
\subfigure[Parallel efficiency.]{
\includegraphics[scale=0.31]{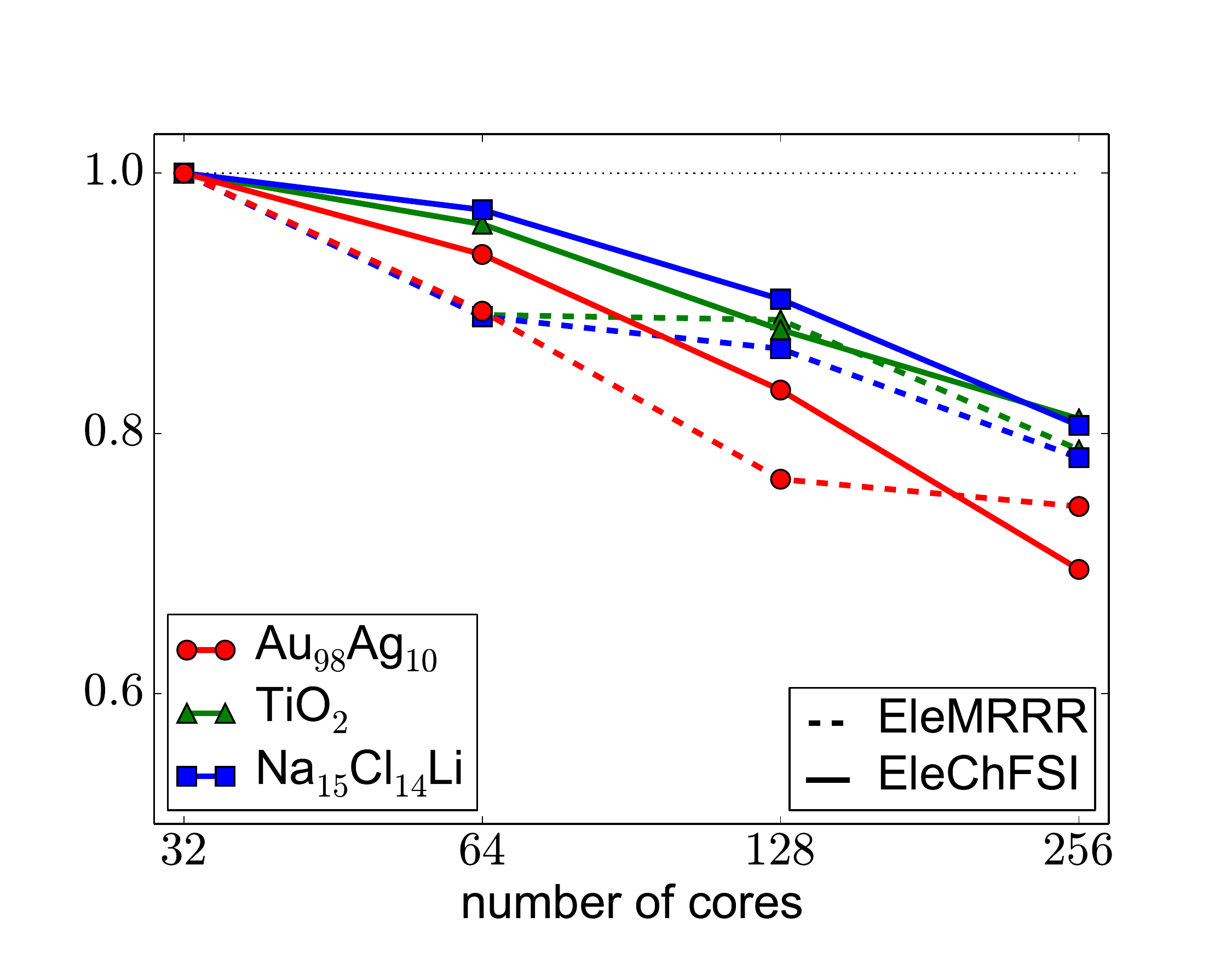}
}
\caption{EleChFSI behaviour for an increasing number of cores. In plot
  (a) it is depicted the strong scalability where the size of the
  eigenproblems are kept fixed while the number of cores is
  progressively increased. Plot (b) shows the parallel efficiency of
  EleChFSI. In both plots the green lines refer to the smaller among the
  TiO$_2$ systems.}
\label{fig:scala}
\end{figure}

In our first test we fix the size of the eigenproblem and solve it
using an increasing number of cores effectively distributing the
computational load over more processes by decreasing the ratio of data
per process. Such test measures the strong scalability of the
eigensolver and its outcome is shown in plot (a) of
Fig.~\ref{fig:scala}. Here CPU times are plotted on the y-axis using a
logarithmic scale. Each line corresponds to an eigenproblem of
specific size $n$ which corresponds to a distinct physical system and
need to be solved for a different percentage of the eigenspectrum. In
particular Na$_{15}$Cl$_{14}$Li, TiO$_2$ and Au$_{98}$Ag$_{10}$
require respectively \textsc{nev} $=256\ (\sim 2.8\%),\ 648\ (\sim
5.2\%)$, and $972\ (\sim 7.2\%)$. Despite the eigenproblem properties
and spectrum distribution are quite dissimilar they all have very
similar scaling ratios showing how the scalability of EleChFSI is not
impacted by the nature of the problem.  Moreover EleChFSI is extremely
efficient even when the ratio of data per process, at the rightmost
end of the x-axis, is getting quite low and far from optimal.

In our second test we compute the parallel efficiency of EleChFSI and
compare it to the efficiency of a dense linear algebra (direct)
eigensolver. Parallel efficiency is a measure of the loss of
efficiency of an algorithm as the number of cores increases and is
defined as
\[
\eta_{\textrm{strong}} \doteq \frac{t_{{\rm ref}} \cdot p_{{\rm ref}}}{t_p \cdot p},
\]
where $t$ and $p$ indicate the CPU time and the number of processors.
In plot (b) of Fig.~\ref{fig:scala} the parallel efficiency
$\eta_{\textrm{strong}}$ of EleChFSI is graphed for the same three
physical systems of plot (a) with the addition of three other dashed
curves corresponding to the parallel efficiency of the direct solver
EleMRRR solving for the same problems. This plot clearly shows that
our iterative eigensolver has an overall higher parallel efficiency
than the dense counterpart. Such conclusion can be interpreted on the
basis of the following observations. While EleChFSI performs more
floating point operations ({\it flops}) per data, it also relies
heavily on the parallel BLAS 3 {\tt GEMM} routine implemented in
Elemental which can reach a performance close to the theoretical peak
of the CPUs. EleMRRR, as any comparable direct eigensolver, performs a
smaller number of flops per data but executes a great portion of the
computation -- i.e. the reduction to tridiagonal form -- by leveraging
only BLAS 2 kernels which are quite less performant than {\tt GEMM}.

Currently most FLAPW-based codes still rely on the use of ScaLAPACK
eigensolvers. As shown in~\cite{Petschow:2013vc} ScaLAPACK {\sc
  pzheevx} (Bisection and Inverse Iteration) and {\sc pzheevd} (Divide
\& Conquer) routines are less efficient than EleMRRR for a number of
cores higher than respectively 128 and 256 while {\sc pzheevr} (MRRR)
efficiency drops starting at 512 cores. These routines are used to
solve for $\mathcal{N}\, _k$ eigenproblems per SCF using few hundreds of
cores per problem with residual tolerance close to machine
precision. So in the best case scenario each SCF can use efficiently
up to $\mathcal{N}\, _k \times 512$ cores.

The parallel efficiency plot of Fig.~\ref{fig:scala} makes manifest
EleChFSI's potential to scale efficiently on larger number of cores than
the ScaLAPACK routines. The slow decrease of the three solid lines
ensure that, for eigenproblems of size $n > 10-15,000$, EleChFSI can
do a better job in exploiting large distributed architectures. In
other words EleChFSI has the potential of allowing the users of
FLAPW-based codes to generate more complex physical systems made of
thousands of atoms resulting in eigenproblems of sizes reaching up to
and over 100,000. Additionally the ability of EleChFSI to obtain
solutions with a lower accuracy without affecting the convergence of
the sequences implies further savings in computational time.

\subsection{Performance on sequences of eigenproblems} 

\begin{figure}[!htb]
\hspace*{-0.7cm}
\centering
\subfigure[Na$_{15}$Cl$_{14}$Li - $n=9,273$ - {\sc nev} $=256$.]{
\includegraphics[scale=0.31]{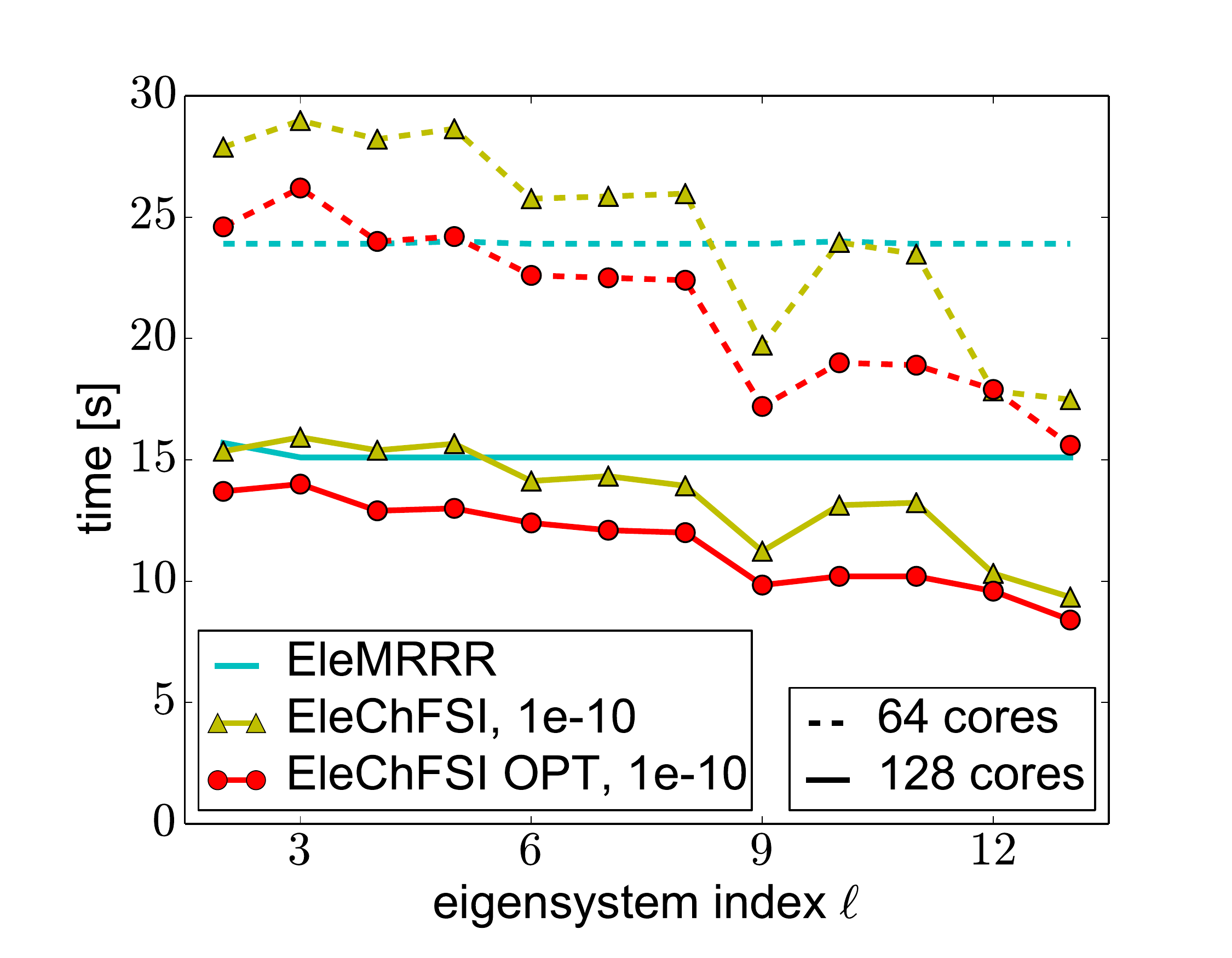}}
\subfigure[TiO$_2$ - $n=12,455$ - {\sc nev} $=648$.]{
\includegraphics[scale=0.31]{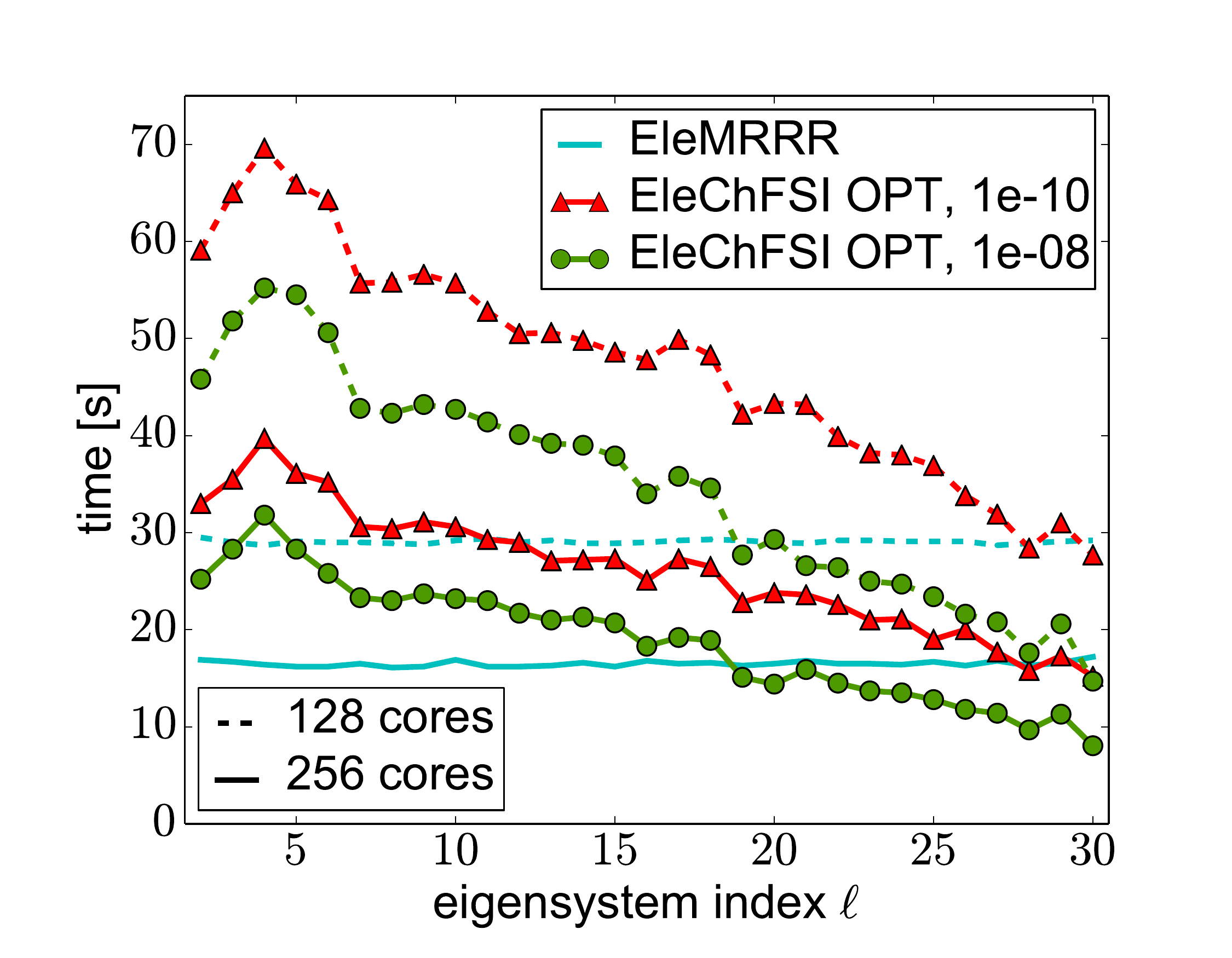}}
\caption{Comparing EleChFSI with EleMRRR on eigenproblems of
  increasing self-consistent cycle index $\ell$. For the size of
  eigenproblems here tested the ScaLAPACK implementation of BXINV is
  comparable with EleMRRR~\cite{Petschow:2013vc}. For this reason a
  direct comparison with the BXINV solver is not included.}
\label{fig:ItvsDir}
\end{figure}

In this subsection we present a set of numerical tests meant to
illustrate the performance of EleChFSI as a function of the sequence
index $\ell$. The foremost objective is to show how the algorithm
harnesses the correlation between adjacent eigenproblems. For this
purpose we have calculated the speed-up as the fraction of CPU times
EleChFSI requires for computing the solutions when it uses initial
random vectors as opposed to when it is inputted approximate
solutions. Plot (b) of Fig.~\ref{fig:timeprof} unequivocally shows
EleChFSI obtains a great advantage from the use of the eigenvectors
$Y^{(\ell-1)}$ as input in solving the next eigenproblem $H^{(\ell)}$
in the sequence. Already at the beginning of the sequence the
algorithm experiences speed-ups higher that 2X and well above 3X
towards the end of it. From the analysis of similar plots for other
sequences we observed this behaviour is stable with respect to small
changes of the sequence correlation and, as such, does not depend on
the physical system or spectral properties of the eigenproblems. In
other words taking advantage of the sequence correlation is a
universal strategy which can be systematically implemented.

Compared to direct solvers, EleChFSI promises to be fairly
competitive. Depending on the number of eigenpairs computed, our
algorithm is on par or even faster than EleMRRR. In plot (a) of
Fig.~\ref{fig:ItvsDir} the non-optimized EleChFSI appears to become
competitive with the direct solver only later in the sequence when
using just 64 cores. The situation improves substantially with 128
cores, and at the beginning of the sequence both algorithms are on
par. This response can be easily explained by observing the higher
parallel efficiency of EleChFSI with respect to EleMRRR efficiency in
plot (b) of Fig.~\ref{fig:scala}. The second set of lines in plot (a)
of Fig.~\ref{fig:ItvsDir} shows the execution time of the single
optimized code. EleChFSI OPT saves an extra $15-20$\% of CPU time and
becomes, in absolute terms, the faster algorithm outperforming EleMRRR
quite dramatically at the end of the sequence. From this point of view
the \emph{Single Optimization} gives EleChFSI that extra edge which
makes the out-performance of dense linear algebra solver possible in
this particular case.

Clearly the above conclusions are not universal and depend on the
fraction of eigenspectrum desired. In plot (b) of
Fig.~\ref{fig:ItvsDir} we observe a different situation where even the
singly optimized EleChFSI does not outperform as easily the direct
eigensolver. In this case one can take advantage of the flexibility of
the algorithm and, if the nature of the application allows it, require
a looser stopping criterion for the eigenpair residuals. In DFT it is
often the case that a threshold of $10^{-08}$ gives already
eigenvectors which are accurate enough for the purpose of the
simulation. With this information in hand EleChFSI remains still
competitive at least for the second half of the sequence.

\begin{figure}[!htb]
\hspace*{-0.9cm}
\centering
  \includegraphics[scale=0.40]{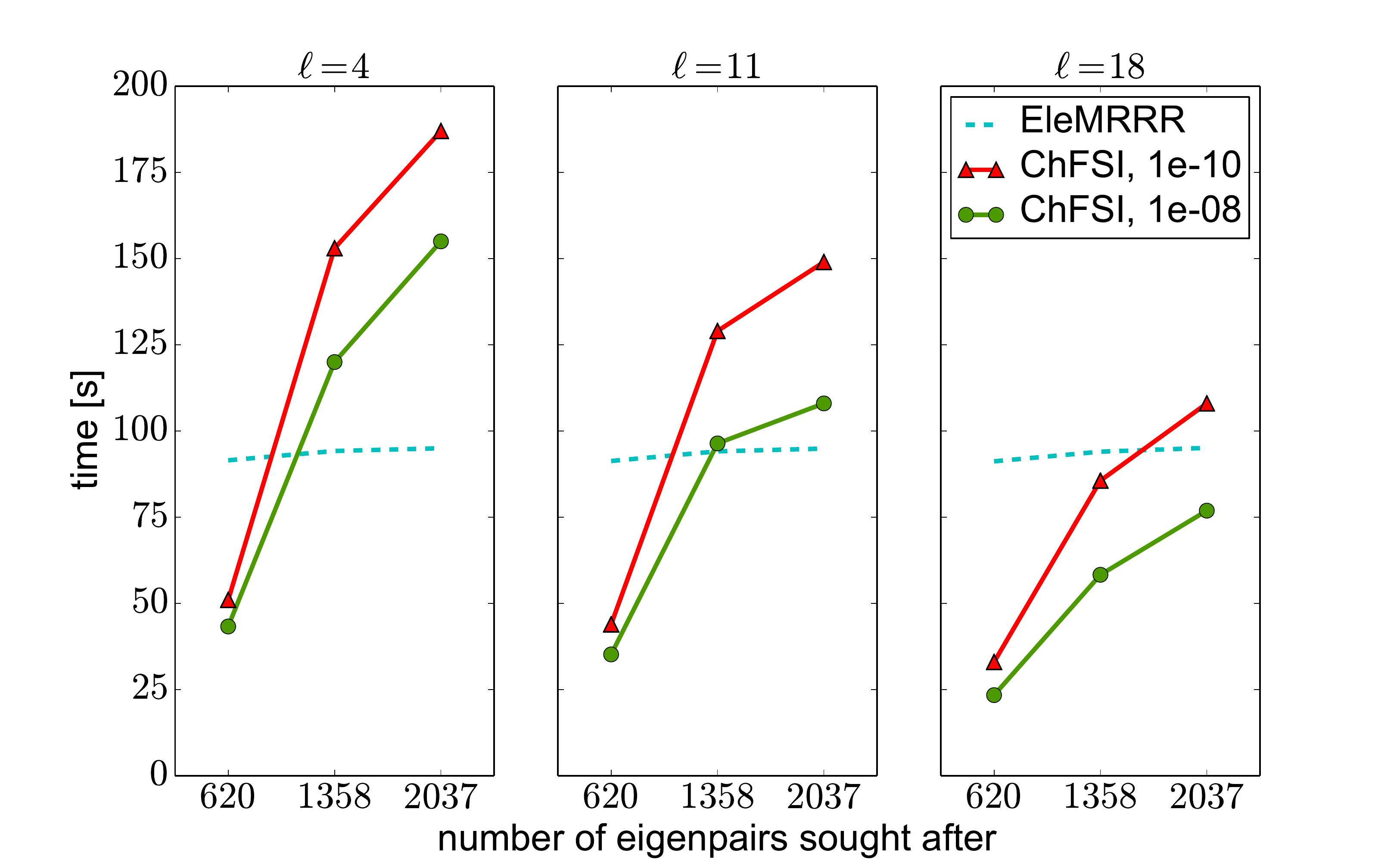}
  \caption{The data in this plot refer to a sequence of eigenproblems
    at $\ell=4,\ 11,\ 18$ of the physical system TiO$_{2}$, with size
    $n=29,528$, and a variable number of sought after eigenpairs
    $\textsc{nev} = 620,\ 1358,\ 2037$, corresponding respectively to
    $\sim 2\%, \sim 4.5\%$ and $\sim 7\%$ of the spectrum. In this
    plot the 3 eigenproblems are solved using EleMRRR and EleChFSI for
    two distinct threshold tolerances, namely $10^{-08}$ and
    $10^{-10}$. }
  \label{fig:fractol}
\end{figure}

In order to illustrate the complex dependence of EleChFSI on the
minimum tolerance for the eigenpair residuals, the eigenspectrum
fraction, and the sequence index $\ell$, we have represented in
Fig.~\ref{fig:fractol} its performance for different values of these
parameters. From this figure it is evident that depending on the
fraction of the spectrum and tolerance desired, EleChFSI can be
competitive for the whole sequence or just part of it. Moreover it is
also evident how the spectrum distribution plays a significant
role. In fact in going from $4.5$\% to $7$\% of eigenpairs the
EleChFSI curves become less steep. This counterintuitive effect is
imputable to the presence of gaps in the spectrum distribution closer
to the end of the filtered interval which the Chebyshev filter is
capable to exploit effectively. It is important to notice that all the
above conclusions are truly independent of the size of the
eigenproblem sequence as it is demonstrated by the relatively large
eigenproblems ($n \sim 10-30,000$) such as the one used in
Fig.~\ref{fig:fractol}.


  \section{Summary and conclusions}
  \label{sec:sumconc}

  In this paper we extend the presentation~\cite{BerDin:LNCS} of the
  Chebyshev Filtered Subspace Iteration (ChFSI) algorithm especially
  tailored for sequences of correlated eigenproblems as they arise,
  for example, in FLAPW-flavoured DFT methods. In particular we have
  illustrated an optimized version of the algorithm and its
  parallelization for distributed memory architecture.  The paper
  introduces the concept of sequence of eigenproblems and
  how the idea of correlation adds critical information to the
  sequence. We show typical examples of eigenproblem sequences as they
  arise in materials science and quantum chemistry and illustrate the
  mathematical model from which they emerge.

  We then describe the algorithm in its details with specific
  reference to the Chebyshev filter. Parallelization is realized
  through the use of the Elemental library framework which hides the
  intricacy of the matrix distribution behind a very efficient
  interface. Numerical tests demonstrate that the parallel
  implementations of ChFSI takes great advantage from the progressive
  collinearity of the eigenvectors along the sequence.  On large
  parallel architectures, the algorithm also proves to be quite
  performant. In the specifics we showed how EleChFSI scales extremely
  well over a range of cores commensurate to the size of the
  eigenproblems. Compared to direct eigensolvers, EleChFSI is
  competitive with routines out of ScaLAPACK and Elemental whenever
  the percentage of the eigenspectrum sought after is not too high.

  The results of this paper paves the road to a new strategical
  approach when dealing with sequences of dense correlated
  eigenproblems. Instead of solving each problem of the sequence in
  isolation, solutions of one problem can be exploited to
  pre-condition an accelerated subspace iteration eigensolver and
  speed-up its performance. This strategy, in the specific case of
  DFT-generated sequences, will enable the final user to access larger
  physical systems which are currently out of reach.

\bibliography{numalg,comphys,mypab,soft}{}
\bibliographystyle{plain}

\end{document}

%% file: symacros.tex
\newcommand\argmin{\mathop\mathrm{argmin}\nolimits}
\newcommand{\odg}[1]{\mathcal{O}(#1)}
\newcommand\be{\begin{equation}}
\newcommand\ee{\end{equation}}
\newcommand\bea{\begin{eqnarray}}
\newcommand\eea{\end{eqnarray}}
\newcommand\bi{\begin{itemize}}
\newcommand\ei{\end{itemize}}
\newcommand\nn{\nonumber}
\newcommand\Ham{\hat{H}}

\newcommand\nr{{\it n}({\bf r})}
\newcommand\kv{{\bf k}}
\newcommand\rv{{\bf r}}

\newcommand\eqalign[1]{%
	\vcenter{%
		\normalbaselines \advance\baselineskip 5pt
		\advance\lineskip 5pt \tabskip=0pt
		\halign{%
			&\hfil $\displaystyle{##{}}$&
			$\displaystyle{{}##}$\hfil\cr
			#1\crcr
			}%
		}%
	}

\newcommand\dotline{\par\hbox to \hsize{\dotfill}\par}

\makeatletter
\def\befored@t#1.#2.#3;{#1}
\def\afterd@t#1.#2.#3;{#2}
\def\refhead#1{\edef\next{\ref{#1}}\expandafter\befored@t\next..;}
\def\reftail#1{\edef\next{\ref{#1}}\expandafter\afterd@t\next..;}
\def\lsim{\mathrel{\mathpalette\@versim<}}
\def\gsim{\mathrel{\mathpalette\@versim>}}
\def\@versim#1#2{\vcenter{\offinterlineskip
        \ialign{$\m@th#1\hfil##\hfil$\crcr#2\crcr\sim\crcr } }}
\newcommand\becomes[1]{\mathchoice{\becomes@\scriptstyle{#1}}
   {\becomes@\scriptstyle{#1}} {\becomes@\scriptscriptstyle{#1}}
   {\becomes@\scriptscriptstyle{#1}}}
\def\becomes@#1#2{\mathrel{\setbox0=\hbox{$\m@th #1{\,#2\,}$}%
        \mathop{\hbox to \wd0 {\rightarrowfill}}\limits_{#2}}}
\makeatother
\newtheorem{formato}{}[section]
\newtheorem{definition}[formato]{Definition}